%% file: main.tex
\documentclass[10pt,leqno]{amsart}

\usepackage{graphicx}

\usepackage{titlesec, xcolor}

\usepackage[obeyDraft]{todonotes}
\usepackage[official]{eurosym}
\usepackage{csquotes}
\usepackage{float}
\usepackage{placeins}

\usepackage[font=small]{caption}
\usepackage{indentfirst,csquotes}
\usepackage[acronym, automake, nonumberlist]{glossaries-extra}

\usepackage[a4paper, margin=2.5cm]{geometry}

\usepackage{amssymb,amsthm,amsmath,array}
\usepackage{xcolor,paralist,hyperref,titlesec,fancyhdr,etoolbox,multirow}

\newcommand{\mssubsection}[2][]{%
  \refstepcounter{subsection}%
  \subsection*{\thesubsection\ #2}%
  \ifx\\#1\\%
  \else%
    \label{#1}%
  \fi%
}

\hypersetup{ colorlinks=true, linkcolor=black, filecolor=black, urlcolor=black }

\usepackage{lipsum}

\makeglossaries

\newacronym{iea}{IEA}{International Energy Agency}

\newacronym{lcoe}{LCOE}{Levelized Cost of Electricity}

\newacronym{egs}{EGS}{Enhanced Geothermal Systems}
\newacronym{orc}{ORC}{Organic Rankine Cycle}
\newacronym{dac}{DAC}{Direct Air Capture}
\newacronym{ocgt}{OCGT}{Open-Cycle Gas Turbine}
\newacronym{ccs}{CCS}{Carbon Capture and Storage}
\newacronym{capex}{CAPEX}{Capital Expenditure}

\begin{document}
\title[risk and reward of a zonal electricity market in Great Britain]{risk and reward of transitioning from a national to a zonal electricity market in Great Britain}
\author[]{Lukas Franken$^{1}$}
\author[]{Andrew Lyden$^{1}$}
\author[]{Daniel Friedrich$^{1}$}
\date{\today}
\address{Address}
\email{lukas.b.franken@gmail.com}
\maketitle

\let\thefootnote\relax
\footnotetext{
$^1$University of Edinburgh, Institute for Energy Systems, Edinburgh, United Kingdom \\
}

\begin{abstract}

More spatially granular electricity wholesale markets promise more efficient operation and better asset siting in highly renewable power systems. Great Britain is considering moving from its current single‑price national wholesale market to a zonal design. Such a reform, however, is argued by stakeholders to entail two forms of risk. First, transitional risk arises in that existing generators located in the north of GB would incur substantial revenue losses as their local wholesale prices reduce relative to a national price. Second, potential increased hedging complexity is argued to weaken the investment case for new generation by raising the cost of capital. This ``permanent" increase of risk could potentially diminish the welfare gains zonal pricing aims to unlock. Existing studies reach varying and difficult-to-reconcile conclusions about the desirability of a zonal market in GB, partly because they rely on models that vary in their transparency and assumptions about future power systems. Using a novel open-source electricity market model, calibrated to match observed network behaviour, this article quantifies consumer savings, unit-level producer surplus impacts, and broader socioeconomic benefits that would have arisen had a six-zone market operated in Great Britain during 2022–2024. 
Regarding transitional risk, in the absence of mitigating policies, it is estimated that during those three years GB consumers would save approximately £9.4/MWh (equalling an average of more than £2.3B per year), but most renewable and nuclear generators in northern regions would experience revenue reductions of 30–40\%. Policy interventions can restore these units’ national market revenues to up to 97\% while still preserving around £3.1/MWh in consumer savings (about £750M per year).
Regarding the ``permanent" increase of risk, it is estimated that the current system could achieve approximately £380–£770 million in annual welfare gain during 2022–2024 through improved operational efficiency alone. The drivers behind these benefits, notably wind curtailment volumes, are expected to become more pronounced towards 2030, suggesting that purely operationally achieved annual benefits of around £1-2 billion beyond 2029 are highly likely. It is found that the scale of these benefits would outweigh the potential downsides related to increases in the cost of capital that have been estimated elsewhere.

\end{abstract} 

\bigskip

\label{sec:intro}

\input{sections/introduction.tex}

\section*{\textbf{The Consumer Perspective on Zonal Pricing}}
\label{sec:results1}
\input{sections/results1.tex}

\section*{\textbf{The Producer Perspective on Zonal Pricing}}
\label{sec:results2}
\input{sections/results2.tex}

\section*{\textbf{Mitigating Transitional Risk}}
\label{sec:transitional_risk}

\input{sections/transitional_risk.tex}

\section*{\textbf{GB Socioeconomic Benefit}}
\label{sec:results3}
\input{sections/results3.tex}

\section*{\textbf{Future Projections of Socioeconomic Benefit}}
\label{sec:future_risk}
\input{sections/future_risk.tex}

\section*{\textbf{Discussion and Conclusions}}
\label{sec:discussion}
\input{sections/discussion.tex}

\section*{\textbf{Methods}}
\renewcommand{\thesubsection}{M.\arabic{subsection}}
\label{sec:methods}

\input{sections/methods.tex}

\section*{\textbf{Acknowledgements}}
The authors would like to express their gratitude to Tim Schittekatte for continued support throughout the making of this paper.
We are grateful to Sam Whitworth, Tom Brown, David Flynn, Andy Hackett, Joe Proffitt and Wei Sun for fruitful discussions and helpful suggestions. 
The authors are particularly thankful to Madelaine Brooks for instigating the project that resulted in this research. 
The authors would like to acknowledge the financial support from Octopus Energy, EPSRC (Engineering and Physical Sciences Research Council) and project partners of the INTEGRATE (EPSRC reference: EP/T023112/1) and DISPATCH (EPSRC reference: EP/V042955/1) projects. 
For the purpose of open access, the authors have applied a Creative Commons Attribution (CC BY) licence to any Author Accepted Manuscript version arising from this submission.

\printglossary[type=\acronymtype]
\addcontentsline{toc}{section}{References}
\bibliographystyle{unsrt}
\bibliography{main}

\section*{\textbf{Appendix}}
\label{appendix}
\input{sections/appendix.tex}


\end{document}

%% file: sections/introduction.tex
\indent The ongoing adoption of intermittent renewable energy sources (RES) puts pressure on electricity markets that were designed around conventional generators like fossil, nuclear and hydro power plants. 
Historically, most power generation was bulky, i.e. hundreds of MWs per installation, located in industrial clusters, and lead times to build those power plants were not that different from the lead time to build out the transmission grid. 
In such a context, coordination between generation and transmission was relatively straightforward. 
Consequently, grid congestion has been rare, at least within the national territories of European countries. 
Therefore, when liberalising since the early 2000s, European countries have commonly opted for a so-called national electricity wholesale market design. 
This means that when clearing the wholesale market, potential violations of thermal transmission limits within a country resulting from the wholesale schedule are not considered, i.e. the national territory equals a ``bidding zone". 
Today, with some exceptions, in the ``coupled" day-ahead and intraday market in the European Union (EU) (referred to as the SDAC and SIDC, respectively), only potential grid congestion between countries is considered. 
The same is true for the electricity market in Great Britain ("GB"), which was ``decoupled" from the SDAC post-Brexit. 
Exceptions in the EU for which the bidding zone does not equal the national territory are Italy (currently consisting of 7 bidding zones), Sweden (4 bidding zones), Norway (5 bidding zones) and Denmark (2 bidding zones) \cite{entsoe2025bzr}.

\indent In contrast to fossil, nuclear and hydro plants, intermittent RES are typically smaller in size, often located in remote areas where little other generation or load has been present, while they take much less time to build. 
Networks cannot be built out as fast as new generation is being built, and the simplification of considering a national territory as a ``copper plate" (as seen by market participants in the wholesale market, i.e. without congestion) no longer resembles reality. 
The most pertinent metric in this regard is the rising redispatch volumes year by year \cite{acer2023mmi, neso2024balancing}. 
Redispatching consists of activations by system operators to ``correct" the wholesale market schedule in a way that it becomes compliant with thermal transmission network limitations. 
This is done by ordering generators (or demand/storage) behind an export constraint to reduce power output (or increase consumption for demand/storage), and units behind an import constraint to do the opposite. 
Under a national wholesale market design, redispatch volumes are expected to rise in the years to come in the EU and GB \cite{neso2024balancing, thomassen2024futureproofing}.

\indent Many European countries that are transitioning to a decarbonised power system are grappling with how to mitigate rising redispatch costs, which are increasingly becoming an important component in the electricity bill. 
Redispatch by itself does not necessarily represent a welfare loss when redispatch mechanisms or markets are seamless. 
However, in reality this is far from trivial. 
Key issues are the potential for gaming (e.g. \cite{stoft1999using}) and the fact that not all assets participate in redispatching processes.

\indent The aspiration to reduce redispatching costs has led regulators and policymakers to think about ways to better reflect transmission constraints in wholesale markets. 
One implementation of such a wholesale market design is Locational Marginal Pricing (LMP), or nodal pricing, which accounts for all transmission constraints in the wholesale market clearing. 
Within North America, all independent system operators (ISOs) other than the Alberta Electricity System Operator (AESO) have nodal pricing in place \cite{westernpowergrid2023electricity}. 
Recently, on the 1st of May 2025, nodal pricing replaced national pricing in Ontario \cite{yen2025ontario}. 
Other large US ISOs have had nodal pricing in place for more than two decades \cite{neuhoff2011international}. 
Further, New Zealand, several Latin American countries, and Singapore also have nodal pricing in place \cite{nzem2002nodal, mcrae2019market, hogan2017singapore}. 

\indent An alternative to nodal pricing is the division of the existing national zone into smaller sub-national zones, often referred to as zonal pricing, which is typically considered a more pragmatic approach in the context of the EU and GB. 
When splitting a national market into multiple zones, the zonal electricity price will fall in zones with excess generation (price risk), and the likelihood that a generator will always be scheduled also reduces (volume risk). 
As a consequence of better reflecting thermal network limits, redispatch volumes will also drop. 
The resulting changes in electricity prices between zones are meant to create natural signals for both producers and consumers to site according to the physical needs of the system. 
In the EU, there is an ongoing Bidding Zone Review (BZR) that assesses the socio-economic impacts of sub-dividing former national wholesale markets into multiple zones \cite{eu2019internal, entsoe2025bzr, agora2025lokale}. 
As mentioned above, some countries have been operating multiple bidding zones for decades, in cases such as Norway or Italy driven by the distance between demand centres and regions of exceptional suitability for hydro generation \cite{harreman2024turning, gotilott2024nodal, birkett2021reforming}. 
In the case of Denmark, Japan and Australia, it is low regional interconnectivity that has motivated the use of multiple bidding zones \cite{gotilott2024nodal, harreman2024turning, skippingstone2024japan}. 
A zonal market approach is also being considered by South Korea \cite{kwag2025quantifying}.

\indent GB stands out as a prototypical example of a national market that may transition to a zonal design. 
Redispatch costs in GB exceeded £2bn in 2023 \cite{mann2024batteries} due to the increasing separation between demand centres in the South and wind generation capacity in the North \cite{Joos2018}. 
In July 2022, the government initiated a Review of Electricity Market Arrangements (REMA), which consults with stakeholders about potential reforms to the GB electricity market and targets a conclusion in summer 2025. 
After a first iteration of REMA, the options being considered in GB will be either retaining the current national market or moving to a zonal design \cite{desnz2022rema}. 
The ensuing debate has reflected the common lines of argument held for and against locational pricing.

\indent On the proponent side, long-term modelling shows that improvements in operational efficiency, more strategic siting of generation and demand assets, and more efficient grid reinforcement could increase GB welfare by between £13bn and £30bn over 15 years \cite{fti2023assessment, fti2025impact}. 
Opponents argue that a zonal market could increase market risk, making a transition to a zonal wholesale market not worthwhile or even increase costs. 
From an investor perspective, this risk is made up of two components.

\indent The first component is \textit{transitional risk} and refers to the potential reduction in revenue faced by existing assets. 
Indeed, besides positive welfare impacts, the introduction of zonal pricing is anticipated to lead to an important transfer from producers to consumers, as found in nearly all assessments. 
This reality has featured significantly in regulators' thinking about policy measures to compensate affected generators, referred to as grandfathering - the practice of ensuring that investments would not financially suffer from the introduction of zonal pricing, as it is argued that this market design change could not have been anticipated when the decision to invest was made \cite{desnz2024rema}.

\indent Despite the number of countries that already operate or have transitioned to operating a locational wholesale market, transitional risk remains an elusive quantity. 
Previous research has estimated that transitional risk for existing assets in exporting regions can manifest itself through a reduction in market value for offshore wind \cite{energinet2022offshore}, generation assets in Texas suffering around 2\% lower energy prices \cite{Eicke2022}, or an estimated 4.8\%–7.0\% reduction in revenues for producers in South Korea \cite{kwag2025quantifying}. 
It is generally assumed that policies to prevent undue revenue losses can be put in place \cite{Eicke2022}; however, existing research falls short of quantifying the effect that such policies, or the lack thereof, could have in GB, leaving consumers and generators with substantial uncertainty about what a market reform would mean for them \cite{blyth2025zonal}.

\indent The second component is potentially \textit{permanently increased risk} for investors under a zonal market.  
Stakeholders argue that a zonal market design would inherently lead to more volatile revenue streams and that hedging is more complicated.  
With regard to the latter, a physical or virtual hub would need to be created to pool liquidity, and instruments for hedging basis risk would need to be introduced, such as financial transmission rights (FTRs). 
It is argued that under such a setup, the cost of capital for new RES generation would increase and potentially wipe out the benefits that a market reform might otherwise achieve \cite{lcpdelta2024zonal, afry2025enhanced, blyth2025zonal}. 
However, there is limited empirical evidence that such an alternative hedging setup would induce more risk and/or liquidity concerns; for a discussion, see \cite{Eicke2022}. 
Studies diverge significantly regarding whether the benefits of zonal markets outweigh these risks. 
FTI Consulting finds benefits large enough to offset an increase in the cost of capital \cite{fti2023assessment, fti2025impact}, while AFRY \cite{afry2025enhanced} and LCP Delta \cite{lcpdelta2024zonal} estimate that a potential rise in the cost of capital of 1\% makes a zonal market a net loss.

\indent In summary, existing forward-looking modelling efforts leave significant ambiguity about the true benefit of zonal pricing in Great Britain. 
As a result, producers face uncertainty about the extent of potential losses under a zonal market, and the degree of influence policymakers have to mitigate those losses. 
Moreover, all stakeholders share a broader uncertainty regarding whether higher capital costs could ultimately negate the benefits that a zonal market might otherwise deliver.

\indent This article analyses the effect that a zonal market in GB would have on the current system and extrapolates the resulting socioeconomic benefits into the future. 
The assessment builds on the novel backcasting GBPower electricity market model calibrated on real-world data, which simulates both a national and a counterfactual zonal market for each day of 2022, 2023 and 2024 at half-hourly and unit-level resolution. 
The benefit of the proposed methodology over other forward-looking approaches is that the model can be validated against real-world electricity market outcomes, which adds to the robustness of the analysis. For each day, wholesale and balancing market prices are sourced from grid operators' data APIs. As a first of its kind, network line capacities are calibrated such that the observed and modelled curtailment volumes match.

\indent Using the GBPower model, this research analyses socioeconomic welfare benefits, impacts on consumer electricity costs and, conversely, unit-level impacts on producer surplus from the introduction of a zonal wholesale market. It scopes the potential for policymakers to compensate for revenue shortfalls for producers and assess the remaining consumer savings after such interventions. With regard to the perceived inherently increased risk under a zonal market design, the socioeconomic benefits are compared against increases in capital costs that could offset these gains.

%% file: sections/results1.tex
\begin{figure*}
    \centering
    \makebox[\textwidth]{\includegraphics[width=1.1\textwidth]{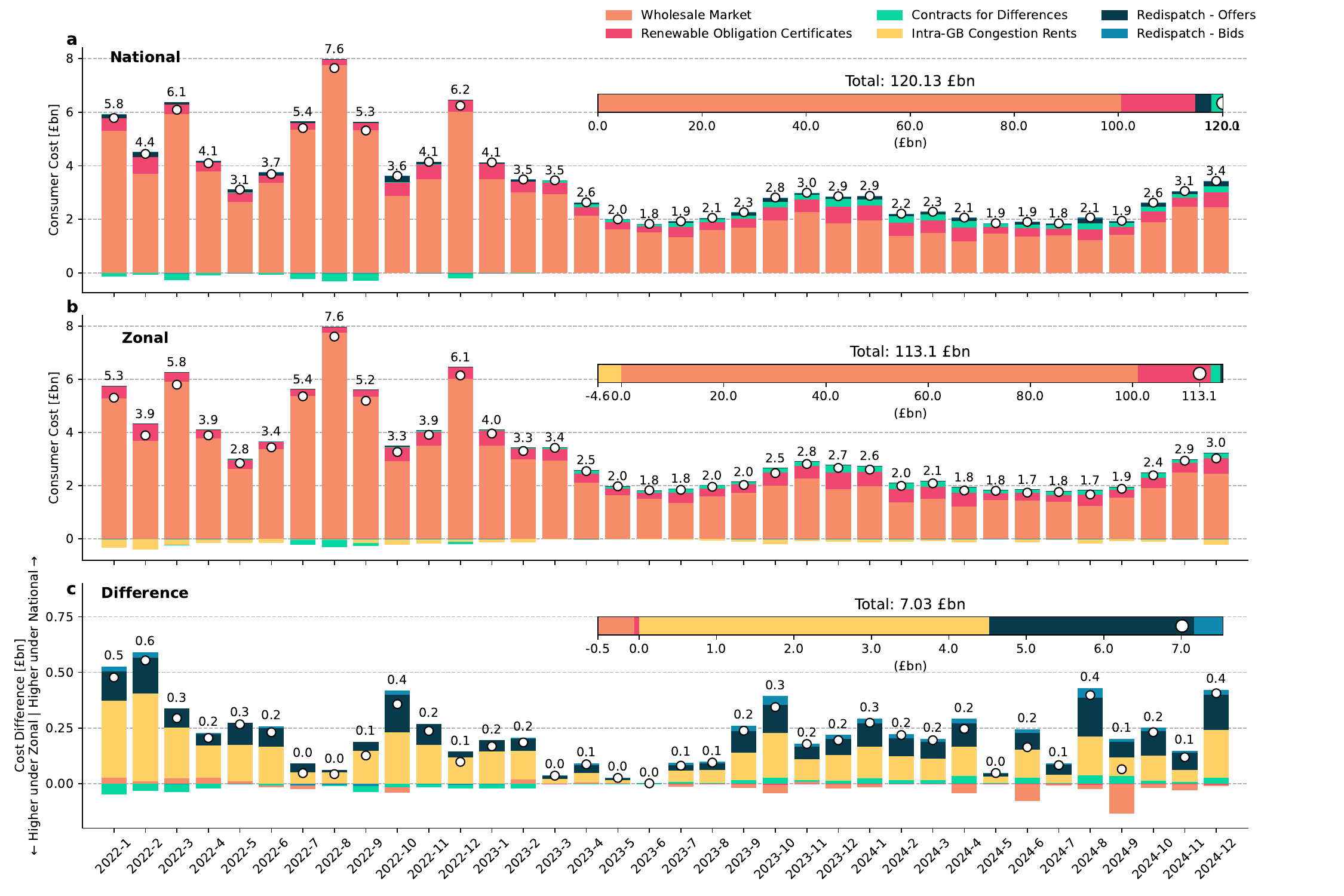}} 
    \caption{\textbf{Total monthly consumer cost under national and zonal market designs, and the resulting differential.} 
Positive differences represent lower consumer costs under a zonal market. 
Redispatch refers only to congestion-related balancing market actions.
The white dot indicates the sum of positive and negative values.
}
    \label{fig:total_monthly_consumer_cost}
\end{figure*}

\indent The GBPower electricity market model is run for each day of 2022, 2023 and 2024, and the outcomes of a national market are compared versus a zonal market with six zones (Fig \ref{fig:layouts}, see Methods \ref{subsec:gbpower}).
For each market design, the model uses daily system load, unit-level generation availability, and observed marginal costs of generation, and then optimises the schedule while accounting for the transmission boundaries visible to the respective market design$^1$\footnote{$^1$For the national market, no transmission boundaries are considered; the zonal market only considers the subset of transmission lines that connect the six zones.}.
The resulting GB wholesale schedules are then assessed against the physical capacity of the transmission grid to evaluate the required volume of redispatching actions for both market designs (see Methods \ref{subsec:balancing}). 
The cost assigned to those redispatch actions, also referred to as ``flagged actions" in GB, via the so-called Balancing Mechanism ("BM")$^2$\footnote{$^2$In contrast to many EU countries, energy balancing actions for frequency support and redispatch actions are performed jointly in the BM rather than in separate markets.}, is based on the observed costs of the actions taken on the relevant day in GB (see Methods for more details). 
In what follows, redispatch costs are referred to as BM costs. 
No impacts of zonal pricing on the need for BM actions to resolve energy imbalances are considered, i.e. so-called ``non-flagged actions" in GB or ``balancing energy activations" in the EU.

\indent GB consumer costs are composed of wholesale market cost, BM costs, inter-zonal congestion rents (which are an income when allocated to consumers), and costs for renewable subsidies, i.e.\ Renewable Obligations (RO) and Contracts for Difference (CfD). 
The analysis focuses on the transmission level and has no representation of distribution-level assets (Methods \ref{subsec:gbpower}).

\indent The zonal electricity market design results in significant consumer benefits, even in the current system. Over the simulated timeframe, when decomposing the different cost components, the national electricity market incurs most of its cost through the wholesale market ($\sim$80\%) and smaller shares through RO payments ($\sim$13\%), net BM costs ($\sim$3.5\%) and CfD payments ($\sim$3.5\%), summing up to a total of around £120bn for 2022-2024 (Fig \ref{fig:total_monthly_consumer_cost} \textbf{a}). 
Under a zonal market, gains in operational efficiency reduce consumer costs on average by $\sim$£3.1/MWh (£2.5bn in total; Fig \ref{fig:total_monthly_consumer_cost} \textbf{c}), indicated by savings in BM costs and a slight increase in wholesale market cost. 
Inter-zonal congestion rents, when allocated to consumers, reduce the consumer cost of electricity by $\sim$£6.3/MWh (£4.5bn in total), resulting in a three-year total cost of around £113bn or a saving of about £7bn over the three modelled years.

\indent The monthly shape of total consumer costs in both market designs is mainly dictated by the fluctuation of fuel (in particular natural gas) prices, caused in large part by the energy crisis \cite{staffell2022power}, and then further superimposed with seasonal variation. 
Consumer costs peak in August 2022 at £7.4bn, coinciding with the height of the energy crisis. 
Lower system load and higher availability of solar generation at the distribution level cause summer months to incur around £1bn lower costs.

\indent The monthly shape of the zonal–national cost differences is mainly driven by wind generation. Excess wind generation causes both higher BM costs in the national market and the accumulation of congestion rent under a zonal market design. 
While wind generation shows no significant seasonal variation, the grid is typically strained during the colder months.  
This is because during winter, the stress of southern wind imports is compounded by higher overall demand and a shortage of solar power supply. 
Further, wind generation capacity gradually expanded over the observed timeframe, causing increased BM costs towards the end of the observed period, showing significant contributions to consumer cost in almost every month between August 2023 and December 2024.

\begin{figure*}
    \centering
    \makebox[\textwidth]{\includegraphics[width=0.8\textwidth]{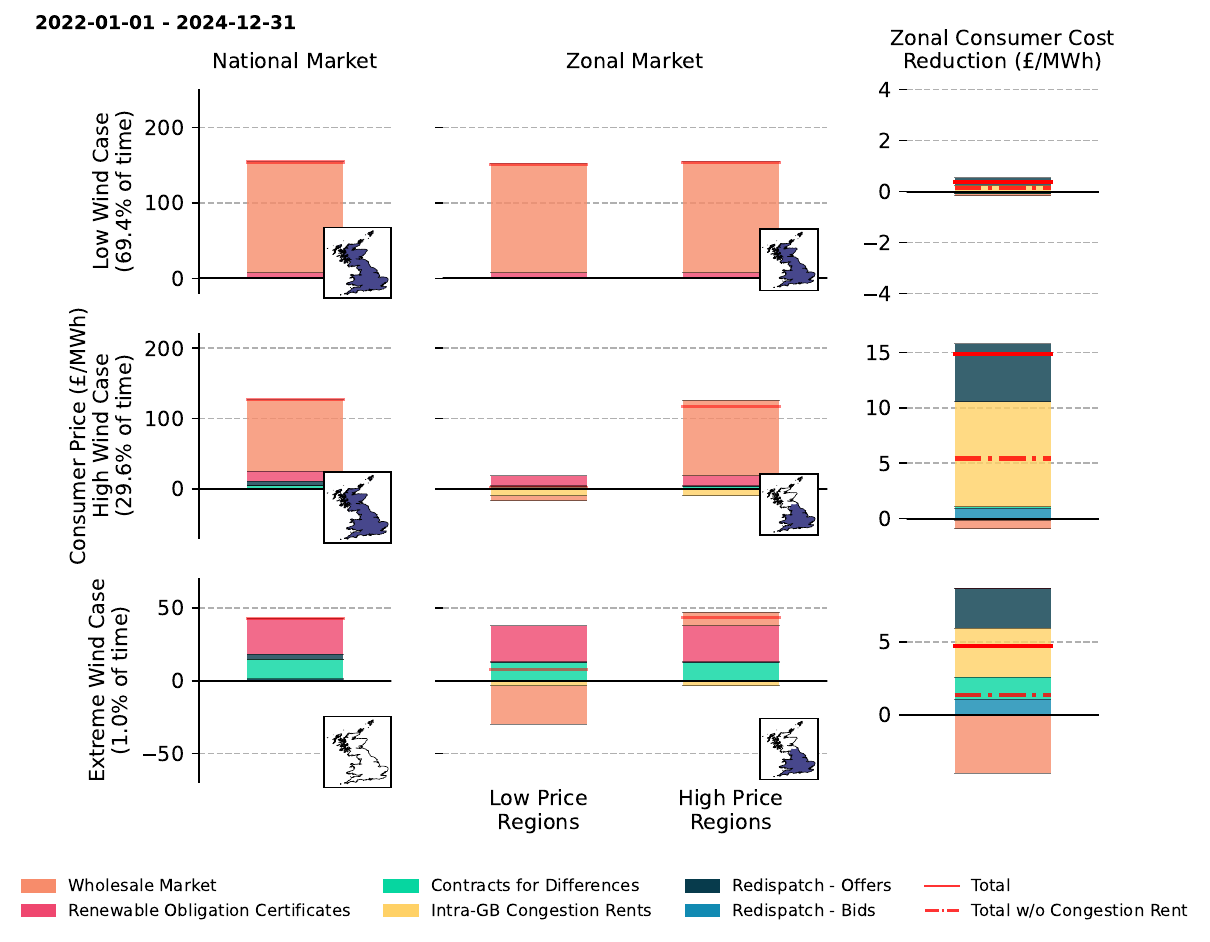}} 
    \caption{\textbf{Electricity consumer cost stack depending on market design and wind availability.} 
    While named after levels of wind availability, settlement periods are classified based on the presence of price-splitting between regions (see small maps). 
    In the third column, negative values represent lower costs under a zonal market. 
    The classification of low- and high-price regions adapts dynamically to the zonal price distribution of the settlement period, and exhibits varying spatial distributions, but a north–south split is most common.
    }
    \label{fig:wind_cases}
\end{figure*}

\indent Delving deeper into the underlying dynamics, the level of wind generation is the main driver behind consumer cost savings from a zonal market. 
In 69\% of simulated hours, wind output is low enough to avoid grid congestion, leading to one (mostly) gas-set marginal wholesale price across GB in a national market and all zones under a zonal market. 
During these periods, a zonal market framework has a negligible impact on consumer costs (low wind in Fig \ref{fig:wind_cases}).

\indent During 30\% of the simulated hours, wind generation is high enough for a zonal market to deliver substantial consumer cost reductions (high wind). 
In those hours, the zonal wholesale prices in the northern zones of GB are typically substantially lower than in southern zones, even though different spatial patterns can occur. 
Wholesale prices in northern zones are typically set by CfD units with (approximately) zero marginal cost, while zonal prices in the South are set by gas-fired generators. 
The resulting large price difference between the zones in those hours leads to high inter-zonal congestion rents, which account for the majority of consumer savings-approximately £10/MWh. 
Additional savings arise from reduced BM costs under zonal pricing, reflecting that a zonal schedule recognises congestion significantly better than a national schedule, adding roughly another £5/MWh of consumer cost reductions.

\indent In cases of extremely high wind availability, which occurs around 1\% of the time, wind sets the price in the national wholesale market.
In a zonal wholesale market, however, transmission constraints may cause the formation of an import-constrained southern zone where gas sets the wholesale price.
This could lead to substantially higher wholesale costs under a zonal market design relative to a national market design in these hours, which could offset the reversed dynamics observed in cases of high (but not extreme) wind availability.
However, at the same time, under a zonal market design, BM costs are substantially lower and high inter-zonal congestion rents are generated. 
The net impact (including congestion rents) is an average £5/MWh electricity cost reduction under a zonal design.

%% file: sections/results2.tex
\begin{figure}[h]
    \centering
    \makebox[\textwidth]{\includegraphics[width=0.8\textwidth]{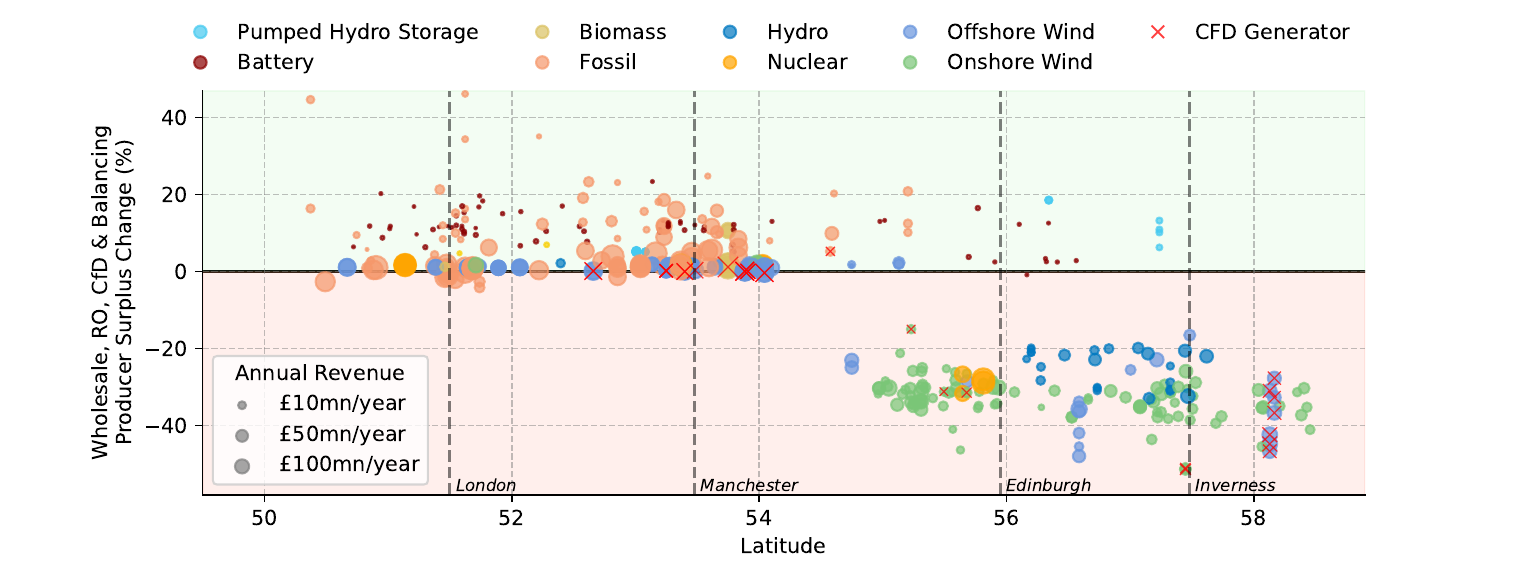}} 
    \caption{\textbf{Percentage total asset-level producer surplus changes over the course of 2022–2024 when moving from a national to a zonal market.}
    For units that are \textit{not} gas-, biomass- or coal-fired, this assumes no marginal costs of generation.
    For thermal units, the changes in surplus are highly sensitive to the assumed premium for balancing services.
    Here, a premium of £30/MWh is assumed for all units providing balancing services; see Fig \ref{fig:surplus_15} for the same figure assuming a premium of £15/MWh.
    In general, assigning balancing volume and cost to specific thermal generators stretches the capabilities of the model.
    The figure does not show northern thermal units PEHE-1, AG-NHAV02, MARK-1, GRMO-1, which see surplus lowered by 72\%, 71\%, 67\% and 63\% respectively.
    The figure does not consider policy intervention to address surplus shortfalls.
}
    \label{fig:unit_revenue_changes}
\end{figure}

\indent Producer surplus would reduce substantially under a zonal market if not addressed through policy. 
For 2022, 2023 and 2024, producer surplus losses on wholesale \& balancing markets and RO \& CfD subsidies are around 30$\pm$5\% for generation units in the North, and there is a small, often positive, impact on the surplus of assets in the South (Fig \ref{fig:unit_revenue_changes}). 

\FloatBarrier

\indent The main drivers of producer surplus impacts appear to be location (though mostly latitude) and technology. 
Wind generators in the North that receive RO payments see large losses averaging around 30\%. 
RO payments are a flat subsidy added per unit of dispatch.
Therefore, these wind generators are fully exposed to the price risk introduced by zonal pricing.
This is different from units operating under CfDs, which guarantee a top-up from any (non-negative) wholesale price$^2$\footnote{$^2$The zonal market assumes that CfD strike prices remain unchanged, and that the local wholesale price is used to determine the CfD top-up.}. 
As such, CfD wind generators are not exposed to the same price risk; however, they still face volume risk as they are less likely to be scheduled in a zonal market due to their unfavourable merit order position relative to RO units (which are assumed to bid negatively, see Methods \ref{subsec:costs}). 
When not scheduled, CfD units receive neither wholesale revenue nor a top-up, and end up with similar losses as RO units.
The magnitude of volume risk for CfD units is surprising for the current system.
However, Fig \ref{fig:wind_cases} suggests that there is (mostly minor) curtailment around 30\% of the time.
With CfD units curtailing first, and wind generation (and therefore wind revenue) concentrated towards times with grid congestion, the value appears plausible.

\indent Nuclear generators in the North have minimal flexibility and therefore are fully exposed to the zonal market's price risk, resulting in losses of around 30\%$^3$\footnote{$^3$To the best of the author's knowledge, among operational nuclear power plants, only Hinkley Point C, in southern England, is supported through a CfD, which could change this outcome.}. 
Northern hydro generators appear to see slightly smaller losses (on average $\sim20\%$), likely because they are dispatchable, which helps to mitigate the financial impact of - on average - lower northern zonal prices. 
In the South, assets experience a slight surplus increase through higher wholesale prices.

\indent Thermal units earn higher producer surplus in the wholesale market (if located in the south where the zonal market clearing prices are elevated) but forfeit a significant portion of their potential revenues from balancing market participation.
Their net change in surplus results from the balance of these two revenue streams.
Balancing market surplus is larger when additional short-term costs (as incurred when balancing on short notice) are low, because then the balancing price premium is driven primarily by strategic bidding, rather than cost recovery, leaving substantial rents for the generator.
Consequently, losing these profits under a zonal design can shrink producer surplus substantially.
Hence, thermal generators' zonal market surplus changes are highly sensitive to assumptions about this additional short-term cost - a quantity that is hard to infer ex ante.
The default value applied here is £30/MWh, consistent with \cite{entsoe2025bzr}, which tips the bespoke balance towards surplus gains for almost all southern thermal units.
When the redispatch cost is halved to £15/MWh (Fig \ref{fig:surplus_15}), average surplus changes for thermal plants are approximately neutral, with unit-specific variations depending on the share of their revenues derived from the balancing market.
Northern thermal units see substantial surplus erosion (beyond 60\% for all of the four northern units included in the model), likely because the zonal market makes them price-setting (i.e.\ eliminating infra-marginal rents) in a substantial share of settlement periods.
Absolute surplus changes for thermal units are shown in Fig \ref{fig:thermal_schedule_revenue}.

\indent Storage units, i.e. batteries and pumped hydro, earn on average about 10\% higher surplus under zonal pricing, independent of their location. 
This is because zonal prices are more volatile than the national wholesale price; 
for 2024 (the year with the least fuel price fluctuation), the national day-ahead price has a standard deviation of £29/MWh, while the zonal market has a standard deviation of around £38/MWh in southern zones, and £58/MWh in northern zones, improving arbitrage opportunities \cite{schmidt2023monetizing}.


%% file: sections/transitional_risk.tex
\begin{table}[h!]
\small
\centering
\caption{\textbf{Analysed policies, implications for northern renewable producers and resulting consumer savings.} Generators not affected by the policies remain at their surplus change from Fig \ref{fig:unit_revenue_changes}.}
\label{tab:policy}
\begin{tabular}{|l|>{\raggedright\arraybackslash}p{4cm}|>{\raggedright\arraybackslash}p{5cm}|>{\raggedright\arraybackslash}p{4cm}|}
\hline
\textbf{Name} & \textbf{Description} & 
\multirow{1}{*}{\begin{tabular}[t]{@{}l@{}}
\textbf{North. Producer Implications} 
\end{tabular}}
& 
\multirow{1}{*}{\begin{tabular}[t]{@{}l@{}}
\textbf{Consumer Savings} 
\end{tabular}}\\ 
\hline
Policy 1 & 
\multirow{3}{*}{\begin{tabular}[t]{@{}l@{}}
Grandfathering CfD strike \\ prices and calculating \\ top-up based on zonal price.
\end{tabular}}
 & \begin{tabular}[t]{@{}l@{}}
Generators 20-40\% (avg 30\%) $\downarrow$.\\
Leaves price risk for RO units.\\
Leaves volume risk for CfD and\\ RO units. \\
\end{tabular}   & £9.4/MWh ($\approx$£2.3bn/a) \\[1ex]
\hline
Policy 2 & 
\begin{tabular}[t]{@{}l@{}}
Policy 1 + \\ 
guaranteed access to pro- \\duction-based FTRs for \\ existing assets. These \\ FTRs top-up wholesale \\ revenues to the weighted \\ average wholesale price \\ (see e.g. Italy's reference \\ price PUN \cite{pun2024}). 
\end{tabular}
 & \begin{tabular}[t]{@{}l@{}}
Generators 0-40\% (avg $\sim$3\%) $\downarrow$.\\
Removes price risk for RO \\ units.\\
Volume risk unchanged.\\
\end{tabular}   & £3.1/MWh ($\approx$£0.8bn/a)
 \\[1ex]
\hline
Policy 3 & 
\begin{tabular}[t]{@{}l@{}}
Split between dispatch \\ and revenue: \\
dispatch based on zonal \\ merit order while \\ revenues are distributed \\
based on counterfactual \\ national market merit order.
\end{tabular}
 & \begin{tabular}[t]{@{}l@{}}
Generators see a uniform revenue \\ reduction of $\sim$3\%. \\
Removes price \& volume risk\\
\end{tabular}   & £3.1/MWh ($\approx$£0.8bn/a)
 \\[1ex]

\hline
\end{tabular}
\end{table}

\begin{figure}[h]
    \centering
    \makebox[\textwidth]{\includegraphics[width=1.\textwidth]{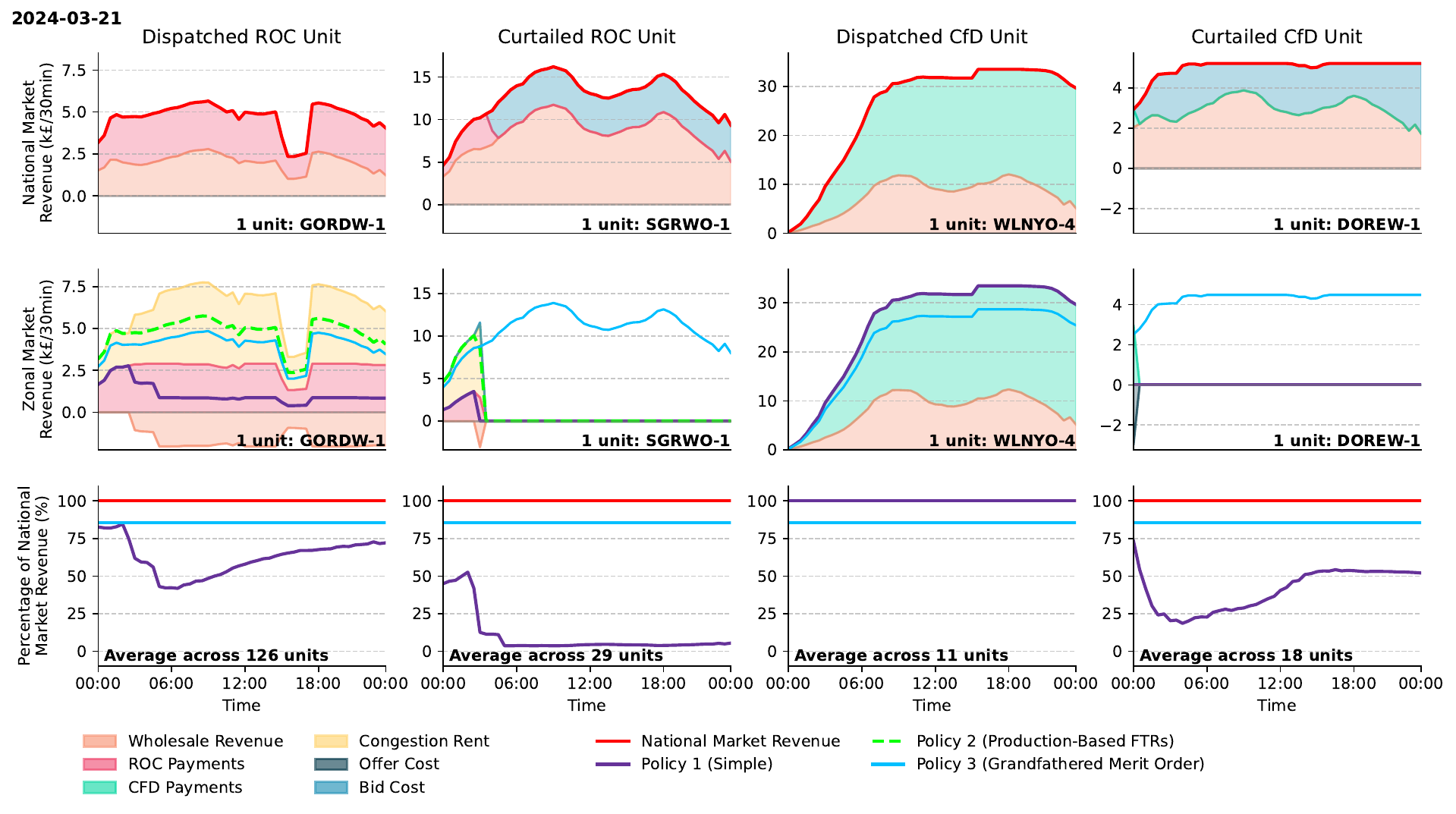}} 
    \caption{\textbf{Revenue stabilisation through the discussed policies for different operational contexts on a congested day.}  
    The plots show one individual unit for different subsidy schemes and balancing market outcomes (dispatched or curtailed) in the national market (\textit{top row}) and zonal market (\textit{centre row}).  
    The \textit{bottom row} shows the total revenues under different subsidies.  
    Units that are partially curtailed are here classified as curtailed, and therefore the average revenue across curtailed units is larger than zero.  
    Policy 2 is redundant for CfD units since they receive a top-up regardless of the sell price (unless the sell price is negative for an extended number of settlement periods \cite{cfd2015contract}).}
    \label{fig:revenue_stabilisation}
\end{figure}

\indent Policies that address revenue losses and thereby transitional risk are at the forefront of regulators' thinking about a transition to a zonal market design in GB.  
As GB's Department for Energy Security and Net Zero (\enquote{DESNZ}) argues, ``\textit{...the past and present investors in GB low-carbon energy are also the investors of the future...}", and further states that managing the transitional risk of existing generators by ensuring reliable revenue streams with minimal uncertainty is a central objective if zonal pricing were to be implemented (Chapter 2 in \cite{desnz2024rema}).  
However, on the side of generators, substantial concerns about how transitional risk shall be treated persists, especially regarding volume risk \cite{blyth2025zonal}.

\indent This section explores policies that could be put in place to ensure financial compensation for the losses suffered by existing generators (Table \ref{tab:policy}).  
\textbf{Policy 1} only grandfathers CfD strike prices, and uses the zonal selling price as the reference for the top-up. This is the grandfathering policy that has been implemented in this paper so far, as it seems largely unquestioned in public discourse.  
\textbf{Policies 2 and 3} pass congestion rent back to northern renewable generators, i.e.\ both reduce zonal consumer savings equally (see Table \ref{tab:policy}; for clarity of exposition, 100\% of rent is passed back here, but the same discussion holds for a smaller share).  
Policies 2 and 3 highlight that different reimbursement mechanisms can yield vastly different outcomes in terms of the distribution of surplus reduction among assets.  
\textbf{Policy 2} eliminates price risk for northern renewable generators.  
The remaining issue - quantified here - is the volume risk a subset of those generators remains exposed to.  
\textbf{Policy 3} tackles that volume risk using a counterfactual national price.  It does so by trimming the price risk top-up slightly, then reallocating those funds to generators exposed to volume risk, striking a compromise between addressing both kinds of risk.
 
\indent In particular, \textbf{Policy 2} uses \textit{production-based} Financial Transmission Rights (FTRs).  
These are rather illustrative and have not seen real implementation, but simplify the discussion and, qualitatively, yield the same outcomes as standard FTRs.  
Standard FTRs are commonly introduced in a zonal market for price hedging between different zones or between a zone and a \enquote{hub} \cite{acer2025virtualhub}.  
In a GB zonal market, FTRs could be used to cover the price (i.e.\ basis) risk between the zone where a unit is located and the zone/hub on which the long-term contract of that asset is indexed.  
Concretely, holders of an FTR obligation are paid out or are required to pay the wholesale price difference between two predefined zones, or between a zone and a hub.  
The actual payment is determined by the hourly price difference and the hourly volume of FTRs held (which is in practice typically flat across a period, e.g.\ month or year).  
Here, production-based FTRs allocate FTR volume proportional to a unit's actual dispatch in the settlement period (making it a perfectly calibrated \textit{production-based} FTR$^1$\footnote{$^1$Production-based FTRs have been discussed in academic literature to address issues of standard FTRs, but have never been implemented due to other issues related to revenue adequacy \cite{ferc2006order681}.}).  
Such a theoretical design would entirely cover the price risk while still highlighting the risk that generators face if they do not dispatch$^2$\footnote{$^2$Production-based FTRs are hard to design and can create undesirable incentives to dispatch at all cost, which could create a downward spiral in wholesale bidding - an issue ignored here.}.  
Policy 2 assumes that these FTRs are granted for free to all parties with assets located in zones that would see a drop in wholesale prices. 
To make sure not more intra-GB congestion rent is allocated to parties than there is collected, the results require appropriate scaling. 
The policy is only applied to generators that are not operating under a CfD.

\textbf{Policy 3} mitigates volume risk by separating units' revenues from their physical dispatch.  
The current national market, de facto, creates revenue streams for renewables that would have not been scheduled in a zonal market - Policy 3 aims to cover that volume risk.  Under the policy, each unit receives an FTR based on the energy it would have been \textit{scheduled} to produce in a national market.  
It thereby shields non-dispatching units from the effects of the market reform by removing the volume risk of Policy 2 and providing payouts to all units proportional to national market revenues.  
That proportion is the share of national market scheduled generation that was actually dispatched, thereby distributing volume risk across assets equally.  
The policy again assumes that total policy payout matches congestion rent volume.  Zonal market dispatch decisions are made based on the zonal merit order curve.

\indent In reality, Policies 2 and 3 are hard to implement without creating bidding distortions.  
What is sketched here is the spectrum of results that are achievable.  
Policy 2, which hedges price risk, can restore national-level revenues for some units while leaving units facing volume risk under-compensated.  
The simulation here estimates the size and residual risk of these groups.  
Policy 3 addresses volume risk by reducing the Policy 2 top-up that units facing price risk receive, and redistributing it equally across assets to cover the volume risk.

\begin{figure}[h]
    \centering
    \makebox[\textwidth]{\includegraphics[width=0.8\textwidth]{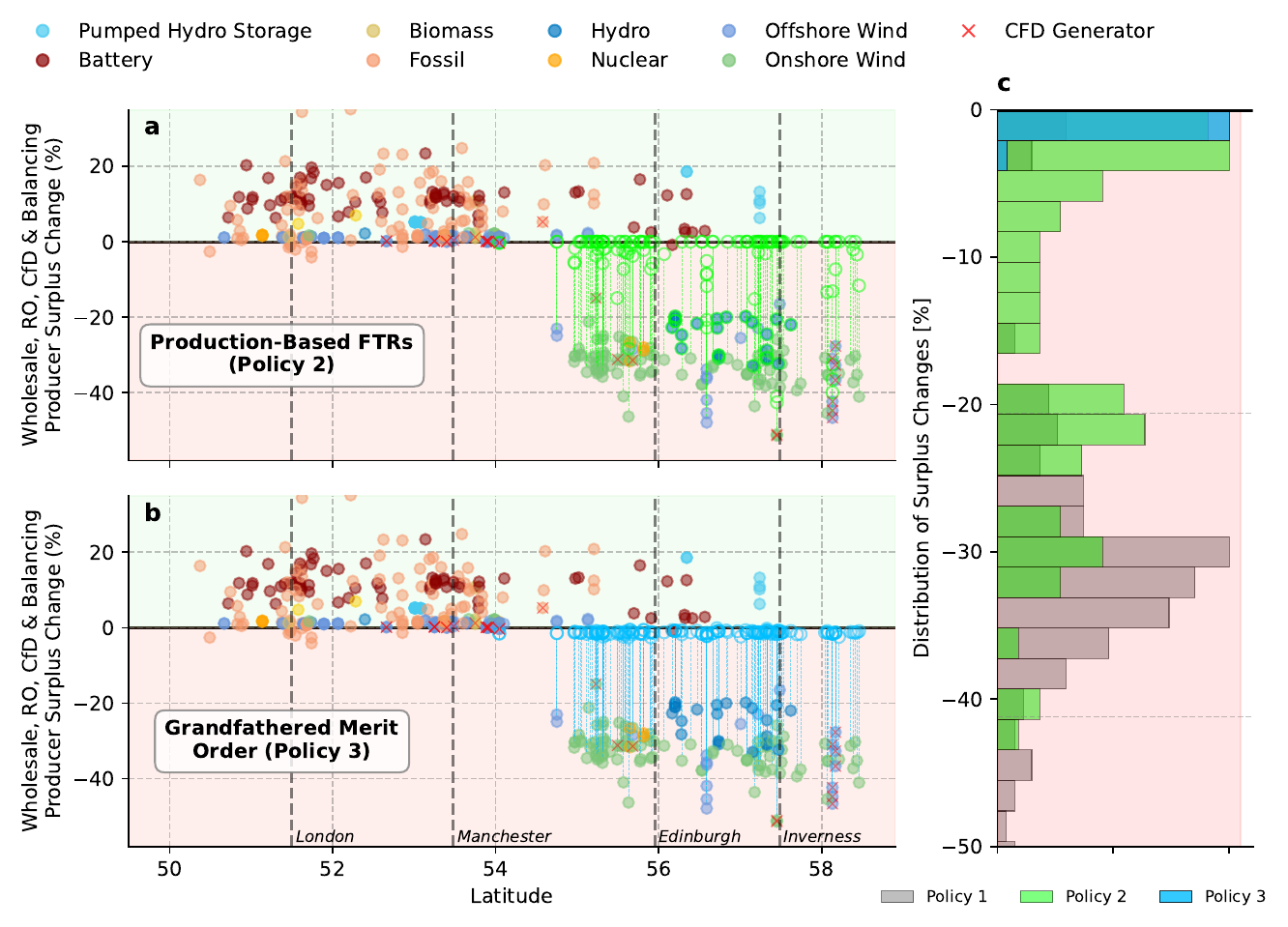}} 
    \caption{\textbf{Producer surplus stabilisation for northern renewable generators under a) Policy 2 and b) Policy 3 over the course of 2022, 2023 and 2024.  
c) Histograms of national market surplus re-instantiation using Policies 1, 2 and 3.}  
The histograms compare the distribution of unaddressed revenue changes (grey) to the distribution of revenues after Policies 2 or 3 are applied.
}
    \label{fig:revenue_stabilisation_annual}
\end{figure}

\indent For an example day - here March 21st 2024, which saw extensive grid congestion - the type of wind generation unit and its dispatch determines the revenues received under the different policies (Fig \ref{fig:revenue_stabilisation}).  
In a zonal market under Policy 1, only the dispatching CfD units receive the same revenue as in a national market (Fig \ref{fig:revenue_stabilisation}, \textit{third column}).  
RO units can achieve profits under a negative local price and therefore are willing to pay for dispatch.  
However, congestion induces negative wholesale prices, and they therefore receive substantially lower revenues than in the national market, on average hovering around 70\% of national market values (Fig \ref{fig:revenue_stabilisation}, \textit{first column}).  
For both of these cases, full national market revenues are restored by production-based FTRs (Policy 2).  
However, the national redispatch market ensures that curtailed units receive the same revenue as they would have if they had dispatched.  
Policy 2 does not provide an analogous mechanism if units curtail (Fig \ref{fig:revenue_stabilisation}, \textit{second and fourth column}), highlighting the volume risk remaining under Policy 2.  
Considering the entire modelled timespan, production-based FTRs allow around half of generators to return to their national market revenues, but leave a substantial share of generators with revenues lowered by around 20\% (Fig \ref{fig:revenue_stabilisation_annual}, \textit{top} and \textit{right}).

\indent Policy 3 rebuilds the majority of revenues that assets in low-price, export-constrained zones lose under a zonal market by leveraging available congestion rent.  
For March 21st 2024, Policy 3 could have restored around 85\% of national market revenues across generators (Fig \ref{fig:revenue_stabilisation}).  
The ``missing" 15\% reflects the generator output that the national market curtailed - but still paid for through the balancing market.  
No corresponding mechanism exists in the zonal market, where the output is never scheduled.  
Therefore, it earns no payment, and congestion rents cannot close the resulting revenue gap.  
Policy 3 distributes this shortfall across all covered generators.  
Over the three simulated years, Policy 3 is able to restore approximately 97\% of national market revenues to all covered generators (because, on average, daily curtailment is substantially lower than on March 21st 2024) (Fig \ref{fig:revenue_stabilisation_annual}, \textit{bottom} and \textit{right}).

\indent Policymakers have expressed interest in both reimbursing generators and passing congestion rents on to consumers.  
The 97\% re-instantiation of national market revenues therefore represents an extreme scenario of regulators choosing to allocate 100\% of rents to generators, with any qualitatively similar outcome achievable under a more consumer-favoured allocation.  Here it is found that allocating the income from intra-GB congestion rents according to our Policy 3, i.e. covering partially for volume and price risk is the most equitable approach. If policymakers seek to exceed a 97\% reimbursement, other mechanisms - such as a consumer tax or a windfall tax on profiting units such as storage - could be implemented.

%% file: sections/results3.tex
\begin{figure}[h]
    \centering
    \makebox[\textwidth]{\includegraphics[width=1.1\textwidth]{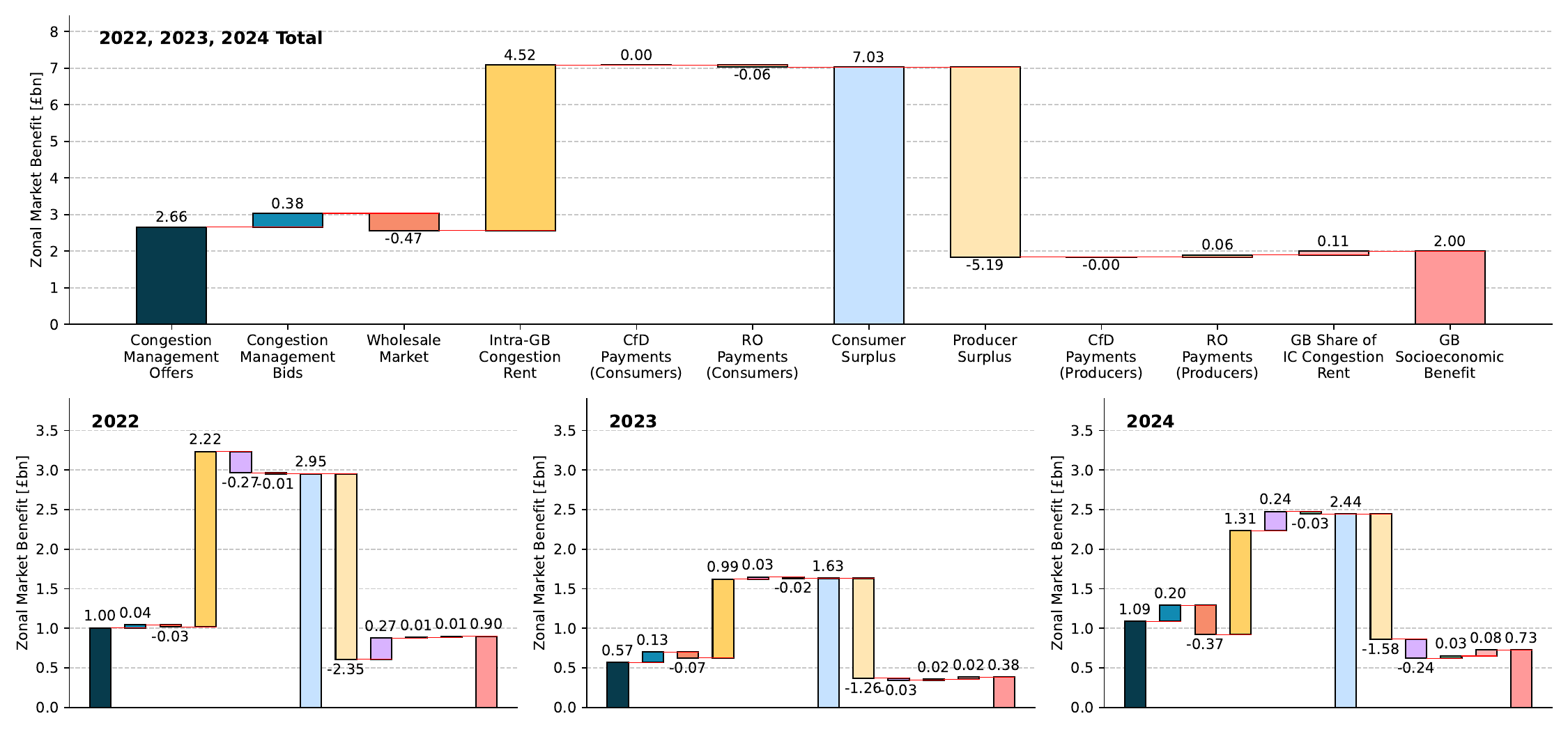}}
    \caption{\textbf{Top-down estimation of consumer surplus, producer surplus and resulting GB socioeconomic benefit / welfare gain.}
    The assessment combines figures from previous discussions of consumer savings (here as consumer surplus) and zonal-induced changes in producer surplus, added to changes to IC congestion rent.
}
    \label{fig:waterfall}
\end{figure}

\begin{figure}[h]
    \centering
    \makebox[\textwidth]{\includegraphics[width=1.1\textwidth]{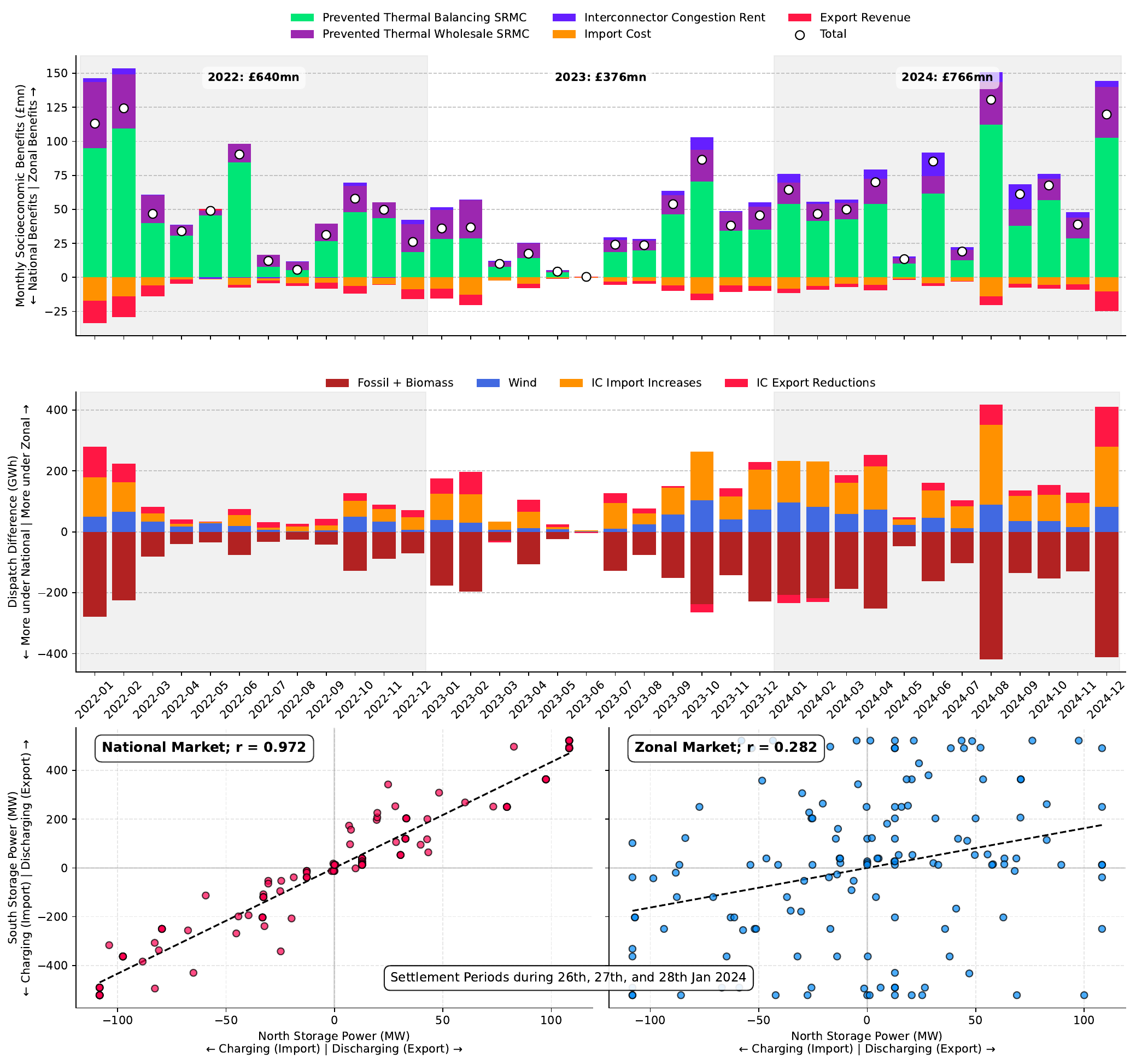}}
    \caption{\textbf{\textit{Top row:} Bottom-up estimation of socioeconomic benefit/welfare gain induced by a zonal market; \textit{middle row:} underlying differences in dispatch; and \textit{bottom row:} correlation between wholesale market position between northern and southern storage units.} 
Positive values in the difference in dispatch refer to higher system usage in the zonal market. 
For details on the components of socioeconomic benefit, see Methods \ref{subsec:seb}. 
Each dot in the \textit{bottom row} refers to the cumulative position across all battery and pumped hydro units in the system's north or south. 
The north–south split follows Fig \ref{fig:north_south_split} and divides assets roughly along the B6 boundary. 
More units are in the south, and therefore their collective charging and discharging capacity is larger.
}
    \label{fig:socioeconomic_benefit}
\end{figure}

\indent Welfare gains, or socioeconomic benefit, arise when the same demand in GB can be served with lower-cost generation. 
Such benefits can arise from having a schedule at the day-ahead stage that better reflects thermal constraints. 
More precisely, under GB's current national wholesale market, wind generation in Scotland frequently depresses the wholesale price. 
The low GB price can prompt ICs, mostly located in the South, to export and storage units in the South to charge, even when north–south transmission bottlenecks are congested. 
This means that in those hours, the schedule of ICs and batteries increases the redispatch volumes relative to a counterfactual in which they would not be present in the system. 
After gate closure, the operator addresses the misalignment in the BM, typically by constraining-off wind in the North and constraining-on thermal generation in the South. 
Reversing ICs in the BM is typically only used as a last resort, and in the years 2022–2024, batteries were not very active in the (congestion managing part of the) BM. 
This means that, because of the suboptimal scheduling of batteries and ICs under national pricing, more thermal generation is eventually dispatched compared to the situation in which the ICs and batteries would have been optimally scheduled at the DA stage. 
This is what happens under zonal pricing, i.e.\ in the same hours, the southern ICs and batteries would have typically been directly “optimally” scheduled in the wholesale market, e.g.\ importing/discharging in the South rather than exporting/charging in hours with significant wind in the North but north–south congestion. 
Better scheduling in the DA market would avoid the need for excessive constraining-on of thermal generation in the BM, which eventually leads to higher cost to serve the same demand.
A zonal market removes most of these misalignments if the zones are well defined. 

\indent One way to quantify the zonal market welfare gain is ``\textit{top-down}" by subtracting producer surplus losses from consumer savings and adding gains in interconnector congestion rent \cite{fti2023assessment}.
The result is a three-year total welfare gain of about £2bn (Fig \ref{fig:waterfall}).
Disaggregated, 2023 contributed markedly less than a third of that ($\sim$£300mn), suggesting that fuels prices (highest in 2022) and curtailment (highest in 2024) drive most of the benefit.
Thermal generator expenses are calculated under the assumption of a £30/MWh balancing mark-up, reflecting genuine technical cost uplift when delivering short-term balancing services.
This value is based on ENTSO-E's assumptions on balancing cost uplift \cite{entsoe2025bzr}.
GB thermal generators, however, historically receive on average about £55/MWh of balancing premium (see Fig \ref{fig:dispatch_price_years});
The £25/MWh gap between the two numbers is therefore a surplus in the national market model, that producers lose under the zonal one (implying the £25/MWh are mostly strategic bidding, and based on the general assumption that opportunities for strategic bidding shrink in a zonal market).
If the entire £55/MWh were purely technical, lost producer surplus would shrink and welfare gain would rise.

\indent The other, ``\textit{bottom-up}", method compares the post-balancing dispatch between market designs, and its operational expenditure.
It better illustrates the mechanics underlying the welfare gain, and can validate the \textit{top-down} calculation.
The results in this paper suggest that, via the above-described dynamics, a zonal market would have achieved £640\,million welfare gain in 2022, £376\,million in 2023, and £766\,million in 2024 (Fig \ref{fig:socioeconomic_benefit} \textit{top}). 
The overall benefit is largely the result of prevented thermal balancing dispatch and to a smaller extent prevented thermal wholesale scheduling.
Increased IC congestion rents are a smaller contributing factor.
These outweigh the added import costs and compensate for lower export revenues (see Methods \ref{subsec:seb} for an overview of techno-economic assumptions).
The annual estimated welfare gain closely matches the \textit{top-down} results.
While 2023 and 2024 track closely, 2022 diverges by around 30\%, likely because more erratic fuel prices complicate the calculation.

\indent The dynamic underlying these welfare gains is a shift in the generation mix, enabled by ICs and storage units;  
increased domestic wind generation, increased IC imports, and decreased IC exports approximately displace dispatched thermal generation under national pricing (Fig \ref{fig:socioeconomic_benefit} \textit{centre}).  
The total volume of unlocked wind over three years accumulated to 1.37\,TWh which - assuming the displacement of gas generation - prevents around 0.24\,MT of carbon emissions.  
The residual imbalance highlights a third contributing factor.  
The zonal market prevents undesirable exports where, in effect, GB exports electricity at a lower wholesale price than the often high prices that must be paid in the BM to make the export possible.

\indent With regard to ICs, the zonal market alters flows during hours of abundant wind. When wind generation is high or extreme (around 31\% of settlement periods; see Fig \ref{fig:wind_cases} \textit{centre and bottom row}), national wholesale prices are depressed, but north–south congestion drives southern zonal prices higher.  
Across 2022–2024, the model shows that in about 10\% of the hours, at least one IC is scheduled in the opposite direction when moving from a national to a zonal market.  
In doing so, ICs are enabled to alleviate local electricity scarcity, revealed through zonal pricing before gate closure time, and contribute to overall welfare gains.

\indent The better scheduling of storage units is a second driver for the reduction in dispatched thermal generation and additional wind energy that zonal pricing unlocks. 
For example during 26–28 January 2024, the model dispatches an additional 11–13\,GWh of wind power per day relative to a national market - the three largest single-day gains across the three modelled years. 
On the same days, the two market designs induce distinctly different charging and discharging patterns of storage units, indicating their influence on the higher dispatch volumes from renewables.  
Under a national price, northern and southern storage units follow the same charge–discharge cycle, shaped by the daily load curve (as indicated by the high correlation in Fig \ref{fig:socioeconomic_benefit} \textit{bottom}).  
This correlation breaks down under a zonal market.  
Northern batteries now respond to zonal price signals reflecting local conditions, and are enabled to schedule around congestion, freeing up network capacity for wind generation to flow south rather than reducing the available capacity.  
The model assumes that each storage unit dedicates around 25\% of its (energy and power) capacity to arbitrage (see Methods \ref{subsec:capacities}); varying this share would magnify or dampen this effect.

%% file: sections/future_risk.tex
\begin{figure}[h]
    \centering
    \makebox[\textwidth]{\includegraphics[width=0.7\textwidth]{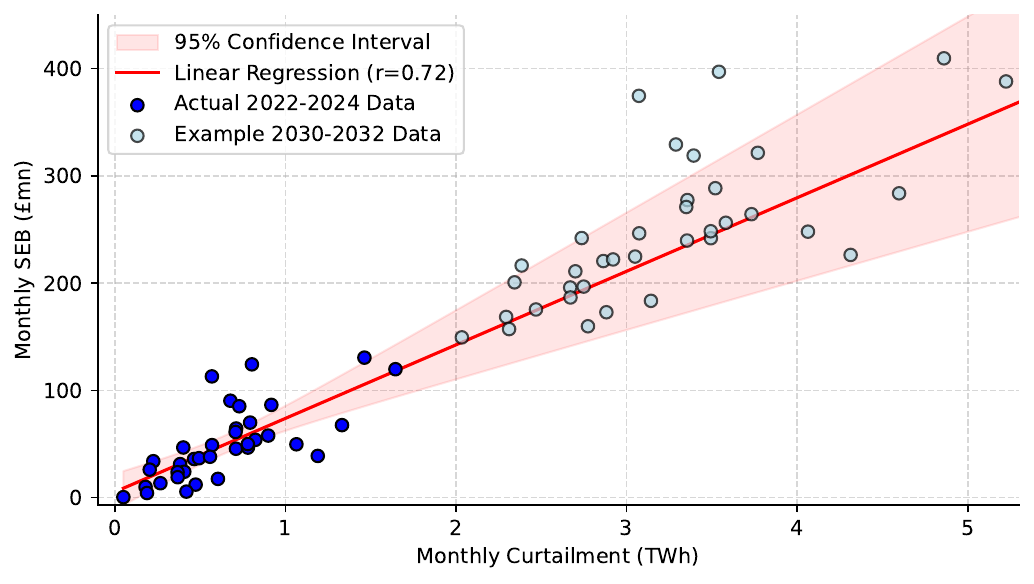}} 
    \caption{\textbf{Monthly correlation between curtailment and zonal market socioeconomic benefit for modelled months, and hypothetical future benefit implications for a 2030–2032 system with higher curtailments.}  
The projections for future curtailment volumes are the same values as for 2022–2024 but scaled to match projections for 2030 and beyond, as estimated by Baringa \cite{baringa2024constraints}.}
    \label{fig:curtailment_seb}
\end{figure}

\indent Socioeconomic benefit is the main metric to evaluate an electricity market reform in GB \cite{fti2023assessment, lcpdelta2024zonal, afry2025enhanced}.  
While most studies agree that a zonal market improves operational - and, in a future system, also siting - efficiency, it remains unclear whether these benefits can outweigh the reform's potential drawbacks.  
This section, building on the modelling of 2022, 2023 and 2024, considers arguments as to why the observed changes in GB welfare may shrink or grow in the future (but does not build on additional modelling).

\indent One known concern voiced by stakeholders is that a zonal market in GB could dent investor confidence, and thereby wipe out achieved welfare gains by driving up the cost of capital \cite{lcpdelta2024zonal, afry2025enhanced}.  
These concerns are typically quantified as \textit{hurdle rate} uplifts - the extra return investors demand before investing. LCP Delta finds that an increase of 0.6\% in hurdle rates erases the benefits of a zonal market \cite{lcpdelta2024zonal}, while AFRY estimates a market reform breaks even at a 0.31\% hurdle rate increase \cite{afry2025enhanced}.  AFRY argues this risk stems from CfD units' merchant tail (with exposure to market prices) during project years 16 to 30, beyond the runtime of their CfD contracts, where the unit is expected to generate half of its lifetime value.  
Higher CfD strike prices or longer contract terms to ensure returns on investment could both consume the welfare gain of zonal pricing.  
In fact, under a pessimistic 1\% increase in capital cost, AFRY predicts the additional zonal market costs outweigh the welfare gain by £7.1bn \cite{afry2025enhanced}, highlighting policymakers' funding gap to adjust subsidies.

\indent The findings here suggest that operating a zonal market in the existing system would have increased GB welfare by between around £370\,million and £770\,million annually over the course of 2022, 2023 and 2024 (Fig \ref{fig:socioeconomic_benefit}).  The underlying drivers - curtailment, "wrongly" scheduled ICs, and untapped flexibility provided by storage units - are all expected to intensify toward 2030.  Annual curtailment volume, which here explains around half of the socioeconomic benefit's monthly variance (Fig \ref{fig:curtailment_seb}), is projected to be five times higher after 2029 than in 2024 \cite{baringa2024constraints}. 
Wholesale prices, depressed by Scottish wind generation and here responsible for wrongful IC use, are likely to be zero around 50\% of the time by 2030 \cite{lcpdelta2024pmo}.  
Further, Future Energy Scenarios 2024 expects that by 2030, the system storage power capacity will have increased to 20–27\,GW, up from today's level of 5\,GW \cite{neso2024fes, gridbeyond2024netzero} - in the present modelling, it is storage units that are the main driver behind the zonal market's ability to unlock wind dispatch.  
Moreover, demand flexibility adheres to the same logic as storage, and is also highly likely to see much larger uptake towards 2030 \cite{neso2024fes}.  
The same holds for ICs, with the 500\,MW \textit{Greenlink} having launched in April 2025, and around 4\,GW of additional mainland Europe–connecting capacity anticipated to come online before 2031 \cite{ofgem2024capfloor}.
The expected trajectory of carbon credit prices are likely to further compound SEB; the reduced UK emission caps towards 2030 \cite{icap2025ukets} and potentially even the linkage with the European Emission Trading Scheme \cite{hancock2024etslink} would further magnify the societal benefit of preventing thermal generation and thereby emissions.
Finally, the wholesale contributions of thermal generators are projected to decline in the future system.  As a result, upward balancing will increasingly depend on starting additional plants instead of ramping partially loaded ones, potentially increasing balancing costs per MWh upward balancing energy provided by thermal units.

\indent Factors that could diminish welfare gain include regulatory changes that help ICs and batteries respond more effectively to congestion within the national market framework.  For ICs, however, this would depend on agreements with overseas TSOs, while for batteries, the gaming opportunities between the national wholesale market and BM have proven hard to eliminate through incremental policy tweaks alone \cite{mann2024batteries}. The same logic holds of flexible demand that seeks to enter the BM.

\indent Given the relationships observed in this research and the outlook for technology deployment, it appears likely that from 2030 the annual socioeconomic benefit will be around £1-2\,billion.
A simple extrapolation based on expected curtailment levels for instance yields about £2\,billion of welfare gain per year (Fig. \ref{fig:curtailment_seb}).  
This would entail a cumulative net present value welfare gain of more than £20\,billion over 2030–2050, even before accounting for additional gains from better asset siting.

\indent This volume of socioeconomic benefit provides policymakers substantially more headroom to absorb or address a higher cost of capital. 
Where AFRY's central scenario finds a £3.3\,billion present-value welfare gain for 2030–2050 - around £170\,million when discounted at 3.5\% per year in real (inflation-adjusted) terms - our model shows projected annual welfare gains five to ten times larger.  Following the arithmetic demonstrated by AFRY, where every 0.01\% rise in hurdle rates cuts zonal market welfare by about £100\,million NPV, the benefits projected in this work could absorb hurdle rate increases significantly larger than 1\%.

%% file: sections/discussion.tex
\indent This paper quantifies the effect that a zonal market would have had if implemented between 2022--2024 in the GB power system.
The effect is measured in terms of consumer savings, the revenue it would shift away from producers, and the resulting net welfare gain.
Based on these findings, the article considers two concerns with regards to the implementation of zonal pricing:
How much of the consumer savings would remain if regulators were to reimburse incumbent generators for their losses (transitional risk)?
And is the increased investor risk in a zonal market likely to outweigh the societal welfare gains (permanent risk)?

\indent With regard to transitional risk, under a zonal market, the vast majority of surplus lost by northern generators accumulates as congestion rent.
In the consumer-friendly scenario, this rent is allocated to consumers, resulting in around £9.4/MWh lower consumer costs, but leaves mostly northern existing generators with a revenue loss of around 30$\pm$5\%. However, if 100\% of congestion rents are used to reimburse generators, the remaining consumer savings are around £3/MWh, while rebuilding around 97\% of national market renewable generator revenues.
Regulators have signalled both a desire to prevent undue hardship for producers \cite{desnz2024rema} and a preference for passing rents through to consumers \cite{desnz2023locational}, and it remains at their discretion which allocation is judged as \enquote{fair}.

\indent The consumer savings found in this work are more than three times as large as the corresponding societal welfare gain. 
In the literature however, typically only around half of consumer-surplus gain translates into welfare gain, see for instance FTI Consulting's zonal pricing assessment \cite{fti2023assessment}. Beyond more efficient investment (not seen by the present purely operational model) that enhances welfare, three other factors are here likely to elevate intra-GB congestion rent (and therefore consumer surplus) that result from features of the present market and the respective approach to model them.
The first is the energy crisis which hugely inflates infra-marginal rents that northern renewable generators receive under a national design.
When transmission constraints are binding, generators do not receive these rents boosting consumer surplus as a result.
The second are RO-accredited wind farms which in 2022-2024 are the dominant share of the wind fleet.
These can submit negative spot-market offers and still receive their certificate revenue to ensure short-term profits.
Compared to future models, where the floor on spot trading is usually £0/MWh, this widens the north-south spread by typically another $\sim$£70/MWh, further increasing congestion rents.
Third, bottom-up models tend to understate within-day wholesale price volatility \cite{mendes2024euromod}.
The present model calibrates the time series of thermal generator short-run marginal prices explicitly to match real day-ahead prices, better reflecting real-world price-fluctuations than typically employed fundamental models which further increases intra-GB congestion rents.

\indent This article finds that a zonal (six bidding zones) electricity market in GB would create around £500\,mn of welfare gain annually in the current system.  
These benefits are achieved purely through better operational efficiency.
The zonal market moves around 82\% of national market curtailment and respective balancing into the wholesale market, where better planning enables more efficient dispatch.
It further visualises electricity scarcity in southern GB, thereby enabling more efficient IC operation.
Finally, over the course of three years, a zonal market enables wind generators to export around 1.4\,TWh of additional electricity by providing more granular price signals to storage units.

\indent In the international context, Great Britain's results cannot simply be extrapolated to other regions, but ENTSO-E's Bidding Zone Review of mainland European electricity markets (BZR) points to broadly similar findings \cite{entsoe2025bzr}.
In terms of overall annual welfare gain, the results here are larger than for Germany and Luxembourg (\euro250\,mn--\euro340\,mn) in \cite{entsoe2025bzr}, which is likely explained by the, in relative terms, lower curtailment volumes in central Europe.
However, inter-annual variation in weather patterns confounds a direct comparison;
ENTSO-E finds that substituting different historical weather years can change the magnitude of outcomes by a factor of two (Section 6.1.2.3.4 in \cite{entsoe2025bzr}).
This is echoed in the present results; welfare gain varies by a factor of around two between years in which the underlying network infrastructure could have changed only modestly.

\indent In terms of future projects and the welfare benefit, several studies warn that a zonal market could lead to higher investment risk, and the resulting rise in the cost of capital could outweigh the benefits of a market reform \cite{lcpdelta2024zonal, afry2025enhanced}.
This article does not dispute nor affirms the claim potential for higher financing costs under zonal pricing, but rather argues that LCP Delta \cite{lcpdelta2024zonal} and AFRY \cite{afry2025enhanced} very likely underestimate the socioeconomic benefit of zonal pricing, and thereby the headroom that policymakers have to address or absorb the higher cost of uncertainty or its mitigation.
The modelling here suggests that from 2030 onwards that a zonal market could likely deliver annual socioeconomic benefits of around £1-2\,bn.
That size of socioeconomic benefits gives regulators substantially more headroom to deal with the potential downsides of a zonal market.

\indent The modelling is based on the current GB power system, which allows the benchmarking of model outputs against real operating data - an option unavailable to future-system studies.
In particular, the accurate replication of power flow is critical, since it is congestion that drives curtailment volumes, and thereby creates the dynamics that cause zonal market welfare gains.
The novel open-source model used here, GBPower, calibrates the transfer capacity of each boundary on every simulated day such that the model's demand for balancing actions matches those used in the real system.
Further, the dispatch price spread between the wholesale and balancing markets is reflected in the model by inserting real day-ahead market prices into the wholesale simulation, and actual bid/offer data into the balancing simulation (Methods \ref{subsec:costs}).
The model is subject to limitations (Methods \ref{subsec:limitations}), but to the best of the authors' knowledge, it achieves the highest accuracy of any open-source model that simulates different electricity market layouts in GB. While these findings contrast with some forward-looking studies, the open-source model used here is calibrated to the observed behaviour of today's network and matches unit-level export potential, curtailment volumes, and wholesale and balancing prices.
The authors therefore believe it provides a more transparent and empirically grounded picture of the dynamics that govern a zonal electricity market in GB.

%% file: sections/methods.tex
\mssubsection[subsec:gbpower]{\textit{GBPower} model}

The simulation of the GB national electricity market, and the counterfactual six-zonal locational market layout, is run in the open-source linear optimisation framework \textit{Python for Power System Analysis} (PyPSA) \cite{brown2017pypsa}.
The novel electricity market model GBPower aims to emulate past days of the Great British wholesale and balancing markets by inserting unit-level marginal costs of dispatch and availability.
The model uses a nodal (around 300 regions) network layout to compare the national and zonal wholesale schedules against a configuration that aligns with transmission constraints. 
The present research is based on model executions for every day between 2022-01-01 and 2024-12-31, one day at a time with 48 settlement periods (i.e.\ in 30-minute intervals). 
The model is, however, functional outside that span but is constrained by the date range covered by the underlying data APIs, including Elexon \cite{elexon2025api}, ENTSO-E \cite{entsoe2025api}, and NESO \cite{ngeso2024constraintflows}.
The model's code is similar in structure to the European-level energy system model PyPSA-Eur \cite{horsch2018pypsa, neumann2023potential}, but it only uses a small number of scripts and data from the original model.
Instead, it focuses on operation (not investment), inserts historical data, and uses validation to approximate the real system as closely as possible.
The following is a discussion of the components that make up the model.

\begin{figure}[h]
    \centering
    \makebox[\textwidth]{\includegraphics[width=0.5\textwidth]{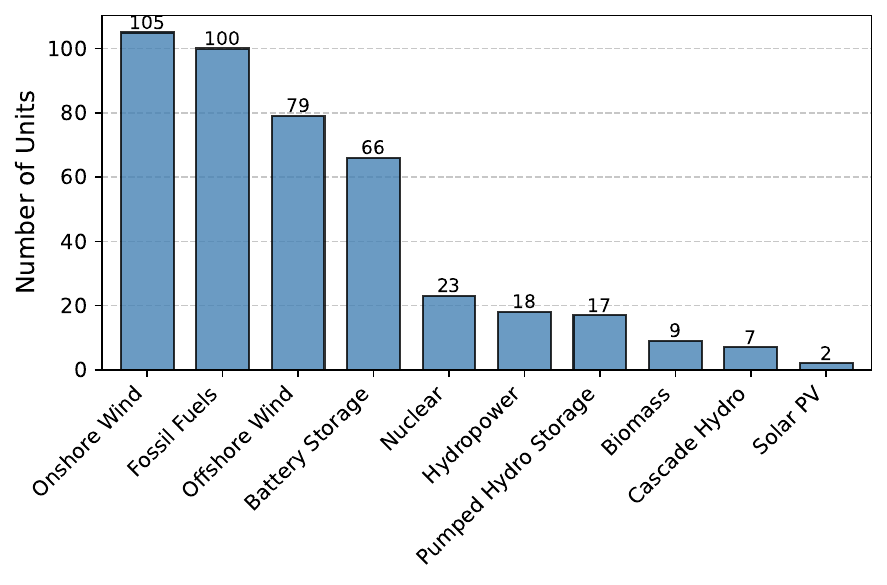}} 
    \caption{Balancing Mechanism Unit count by technology included with location in the model. \textit{Fossil Fuel} groups natural gas, coal, oil, and other fossil-fuel-powered plants.}
    \label{fig:carrier_count}
\end{figure}

\begin{figure}[htb]
  \centering
  \begin{minipage}[b]{0.3\textwidth}
    \centering
    \includegraphics[width=\textwidth]{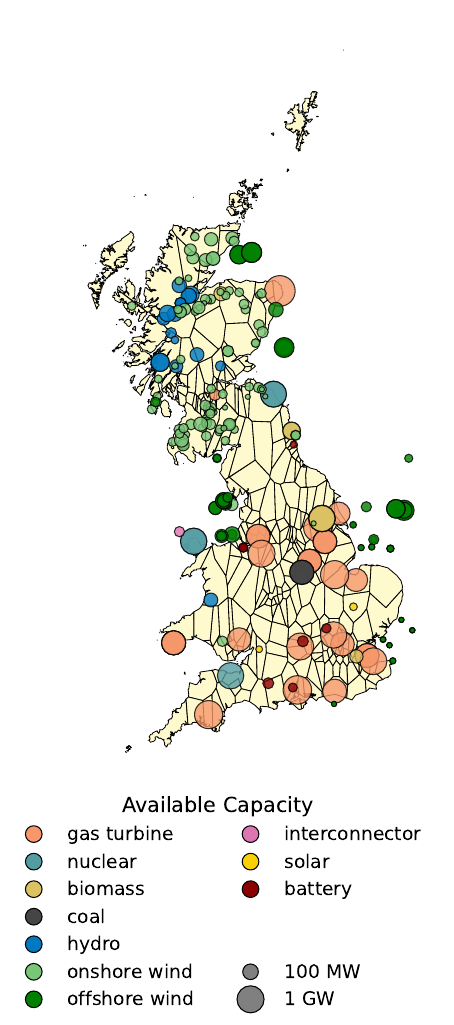} 
  \end{minipage}
  \hfill
  \begin{minipage}[b]{0.3\textwidth}
    \centering
    \includegraphics[width=\textwidth]{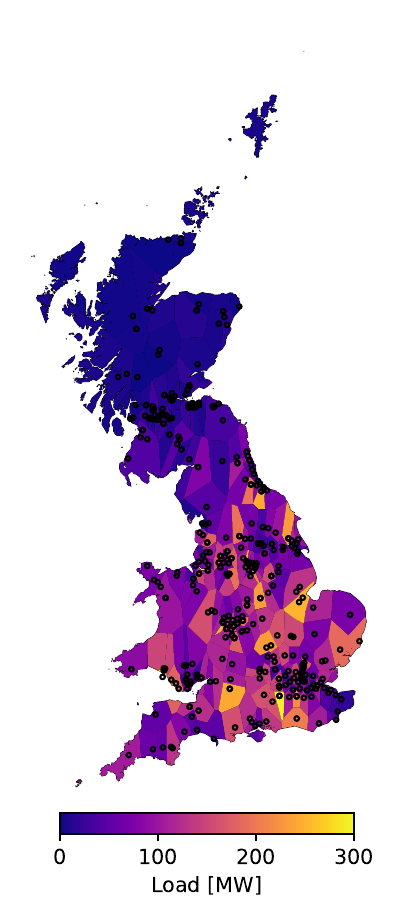} 
  \end{minipage}
  \hfill
  \begin{minipage}[b]{0.3\textwidth}
    \centering
    \includegraphics[width=\textwidth]{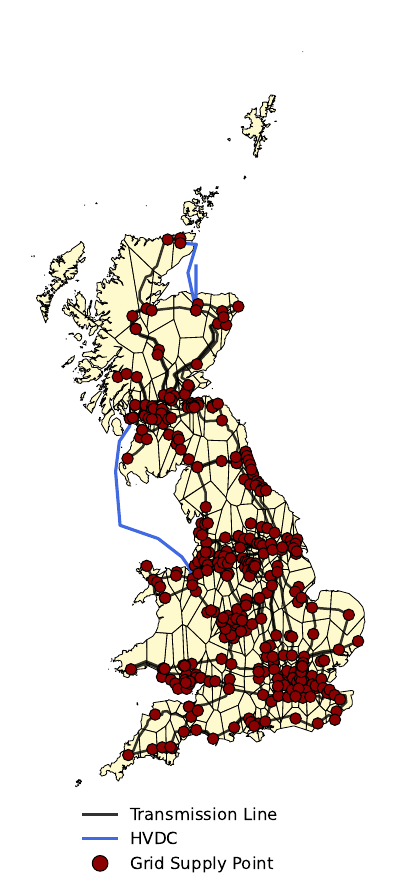} 
  \end{minipage}
  
  \caption{\textit{(Left)} Spatial distribution of generation and storage capacity for an example day, 2024-03-06, and Settlement Period 15. \textit{(Centre)} The respective load distribution. \textit{(Right)} The network buses representing the system's Grid Supply Points and their Voronoi cells, used for the nodal system model.}
  \label{fig:cap_load_network}
\end{figure}

\mssubsection[subsec:locations]{Asset Locations}
The model includes a dataset of 437 individually located generation and storage assets across GB (Fig.~\ref{fig:carrier_count}).
Each of these represents one Balancing Mechanism Unit (BMU), multiple of which can refer to one physical generator.
The model only considers the transmission level and therefore omits units on the distribution level, which drastically reduces the count of some asset classes, in particular solar.
The unit dataset is compiled from the Power-Station-Dictionary \cite{osuked2025bmus} and Wikidata \cite{wikidata2025bmus}, and is complemented with manual searches to locate approximately 99\% of transmission-level units' exports (Fig.~\ref{fig:cap_load_network}).
Export from BMUs that have not been located is distributed (weighted by capacity) among the located ones.

\mssubsection[subsec:markets]{Market Designs}

\begin{figure}[h]
    \centering
    \makebox[\textwidth]{\includegraphics[width=0.7\textwidth]{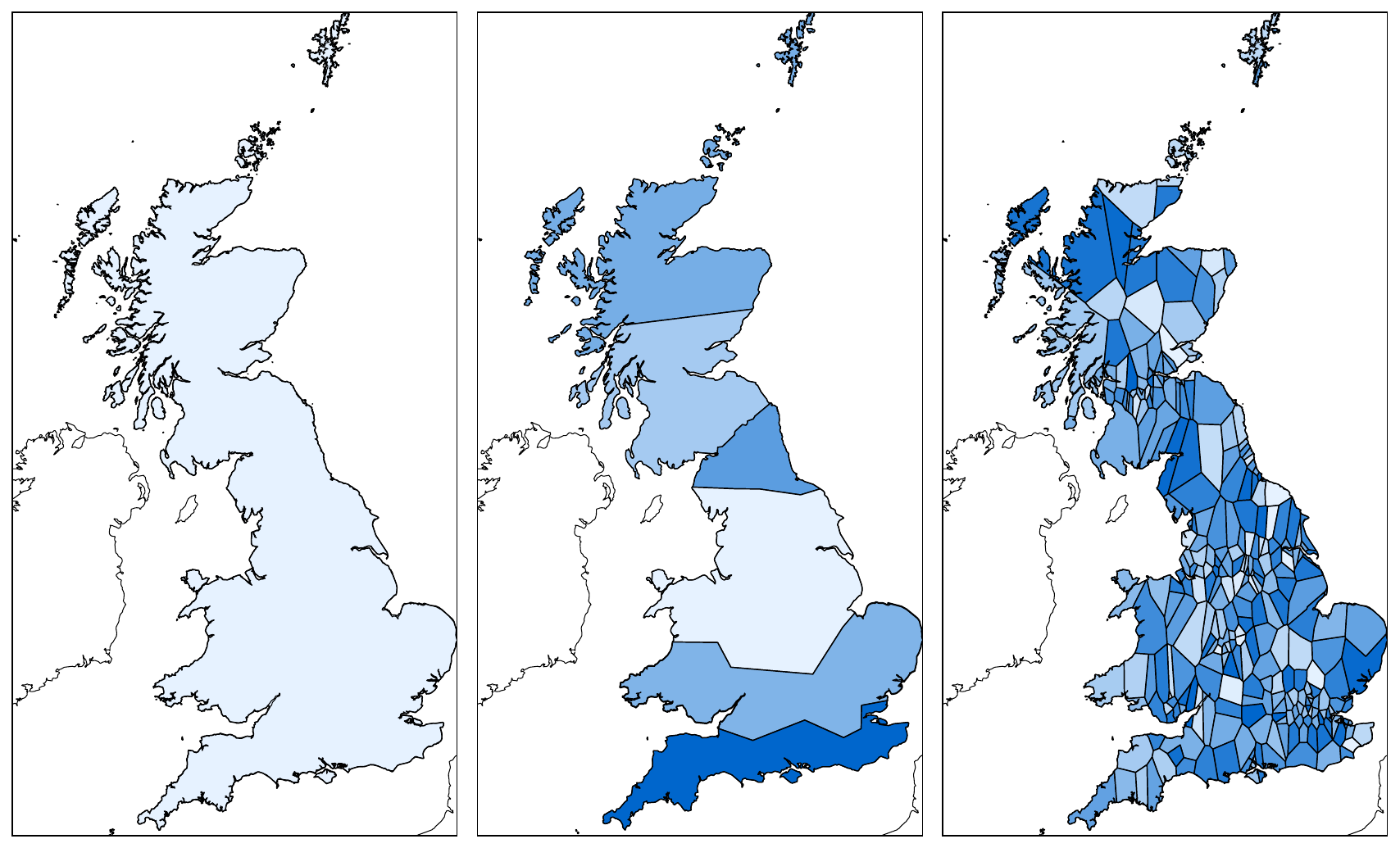}} 
    \caption{Transmission boundaries seen by each of the three market layouts: \textit{national}, (six-) \textit{zonal}, and \textit{nodal}. The nodal layout corresponds approximately to the Grid Supply Points. The boundaries in the zonal layout are chosen to match the grid boundaries for which NESO shares thermal constraints most of the time \cite{ngeso2024constraintflows}.}
    \label{fig:layouts}
\end{figure}

The model includes three market layouts: national, zonal (six zones), and nodal (around 300 zones) (Fig.~\ref{fig:layouts}).
Power flow is simulated using a full DC approximation, as is done in \cite{fti2023assessment}.
This approximation removes constraints imposed by Kirchhoff's law \cite{brown2017pypsa}, and instead simplifies flows across boundaries to a simple Net Transfer Capacity (NTC).
These flows can be compared against real boundaries' thermal constraints, as shared by NESO \cite{neso2015constraints} (see Methods \ref{subsec:network}).
The national and zonal market simulations follow the same procedure.
Daily thermal constraints, wholesale costs, day-ahead prices, interconnector capacity, generation availability, and demand are inserted into the respective layout, and linear optimisation solves for the cost-optimal schedule.
The zonal layout is aware of transmission constraints between zones, but the capacity of all lines that start and end in the same zone is set to infinity.
After wholesale optimisation, the positions of batteries and interconnectors are inserted into a respective copy of the nodal market.
Each copy is optimised for power flow and dispatch to find the transmission-compliant dispatch configuration.
The modelled balancing market is the schedule change that is needed to move from the wholesale schedule to the nodal dispatch.
The model also offers simulation of a nodal wholesale market, which is assumed to already accommodate grid capacities.
As far as this study is concerned, a nodal electricity market has been ruled out in GB \cite{desnz2024rema}, and therefore the findings are not included in this article.

\begin{figure}[h]
    \centering
    \makebox[\textwidth]{\includegraphics[width=0.7\textwidth]{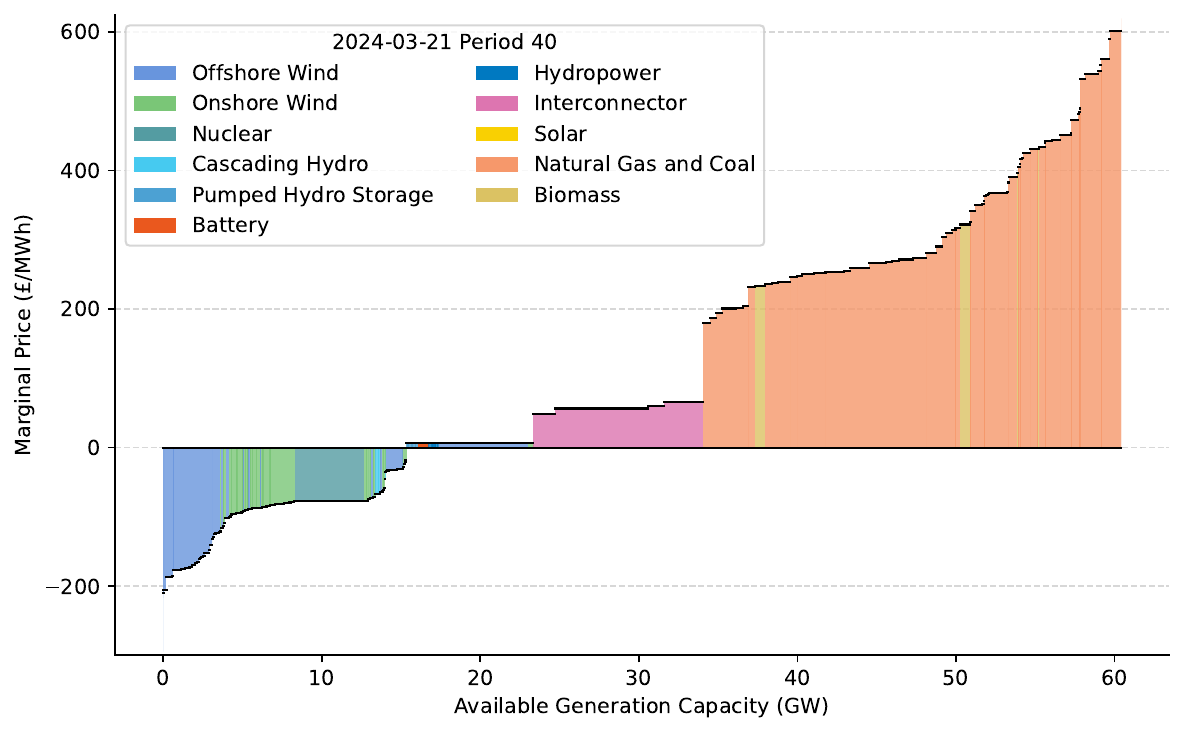}} 
    \caption{Example of the merit order inserted into the model for one settlement period.}
    \label{fig:merit_order}
\end{figure}

\mssubsection[subsec:capacities]{Asset Capacities}
Each unit is assigned a maximum dispatch it can supply in each settlement period.
The heuristics to estimate that number vary between technologies.
There are four qualitatively different types of units: simple generators; daily-quota generators (with flexibility only on when to dispatch a daily quota of available power); storages; and interconnectors - and for each, implementation follows a distinct logic.
Given that the model's goal is to simulate mainly the day-ahead market, only day-ahead Elexon variables are considered.
Costs of generation are discussed in Methods \ref{subsec:costs}.

\vspace{0.1cm} \paragraph{\textbf{Generators}} \hspace{0.3cm}
For wind, nuclear, and solar generators, the respective \textit{Physical Notification} (PN) from Elexon is inserted for every settlement period.
This is because for these units, the \textit{Maximum Export Limit} (MEL) and PNs vary significantly.
This suggests that MEL data reflects a unit's nameplate capacity, rather than its actual availability at any particular time.
Crucially, for thermal units, the model inserts MEL data instead, reflecting that these units have additional (expensive) capacity that can be called upon should renewable resources be insufficient to meet demand (which is almost every settlement period) or should the national or zonal wholesale market require balancing.

\vspace{0.1cm} \paragraph{\textbf{Daily-Quota Generators}} \hspace{0.3cm}
Typical dispatch patterns of hydro (and cascading hydro) suggest that these units have some, but limited, flexibility regarding when to dispatch.
In their physical context, this makes sense;
these generators depend on water streams or run-off for generation (more characteristic of an intermittent unit), but have some, yet limited, capacity to accumulate water while generation is paused.
It appears plausible that such a system would dispatch around periods of high wind availability to maximise market value.
However, it has not been observed that generation is halted for more than a few hours, which suggests that the reservoir size relative to generation capacity necessitates hour-to-hour dispatch decisions.
The inferred ratio of power to energy capacity is relatively small, suggesting that an optimisation horizon of one day can approximate realistic behaviour reasonably well.
In the model, these generators are simulated as storage units that are filled to an initial state (equal to their total actual dispatch for that day), must be fully discharged over the course of the day, and cannot be charged (i.e. import electricity from the grid).

\begin{figure}[h]
    \centering
    \makebox[\textwidth]{\includegraphics[width=0.7\textwidth]{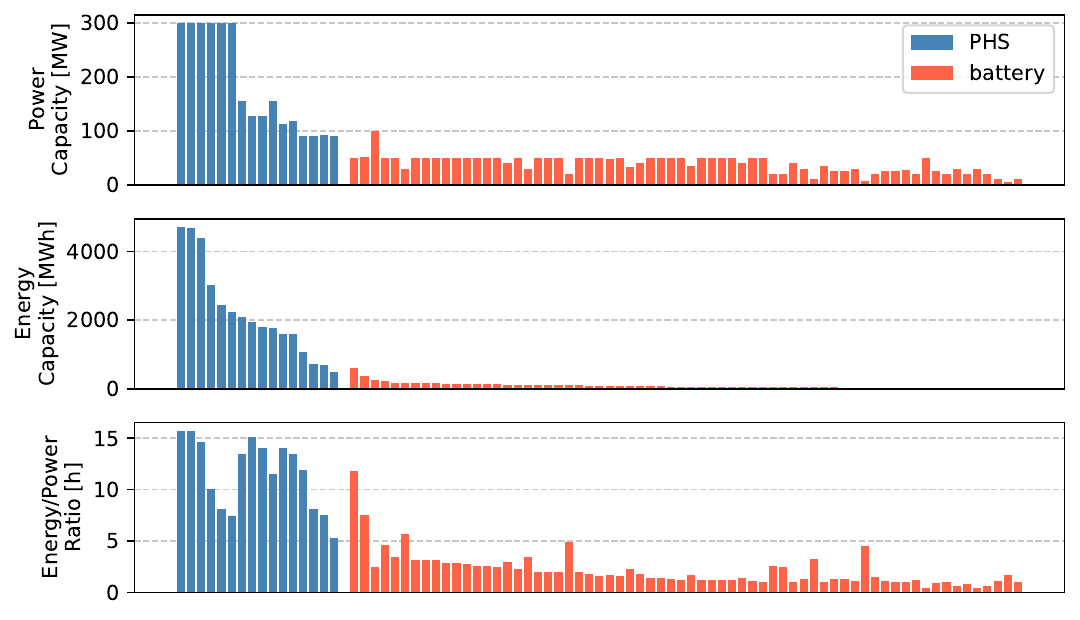}} 
    \caption{Pumped-hydro and battery storages in the model.}
    \label{fig:storage_capacities}
\end{figure}

\vspace{0.1cm} \paragraph{\textbf{Storage Units}} \hspace{0.3cm} 
Pumped-hydro storages and batteries can both import and export electricity from/to the grid.
An assessment of these units' maximum uninterrupted dispatch over the course of 2022, 2023, and 2024 suggests that they operate on a short timescale, engaging in intra-day arbitrage or balancing actions.
For each unit, energy and (dis)charging capacity are here estimated using more than 300 days of PN data.
The model assumes that energy capacity equals the accumulated maximum uninterrupted charging observed within this timeframe, and that power capacity is the maximum charging or discharging observed in any one settlement period (Fig.~\ref{fig:storage_capacities}).
This method recovers, for instance, the power capacity of GB's largest pumped-hydro storage unit, Dinorwig Power Station, represented here as the six largest PHS units (likely representing one turbine each), combining to the plant's nameplate capacity of 1.8\,GW \cite{engie2025dino}.
As a potential limitation, with more than 10 hours of energy-to-power ratio, many of the units might feasibly engage in multi-day arbitrage.
However, actual charging and discharging behaviour over the modelled timeframe did not suggest frequent activity of this kind.
Further, the assignment of multiple BMUs to the same physical unit may distort the data.
During initial stages of testing, the model was enabled to use the full estimated storage capacity in the wholesale market.
However, this led to an overestimation of the true wholesale actions taken by these units.
This is plausible, given that storage units receive a large share of their revenue in various balancing markets.
It was found that wholesale arbitrage behaviour was more realistic once a ``damping" factor was applied, reducing the energy and power capacity of these units (as seen by the model) to 25\%.

\begin{figure}[htb]
  \centering
  \begin{minipage}[b]{0.49\textwidth}
    \centering
    \includegraphics[width=\textwidth]{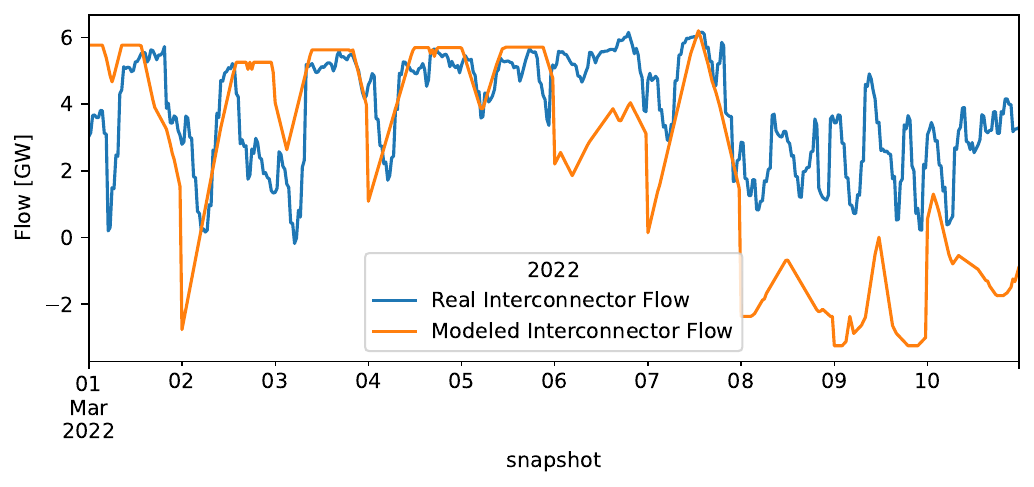} 
  \end{minipage}
  \hfill
  \begin{minipage}[b]{0.49\textwidth}
    \centering
    \includegraphics[width=\textwidth]{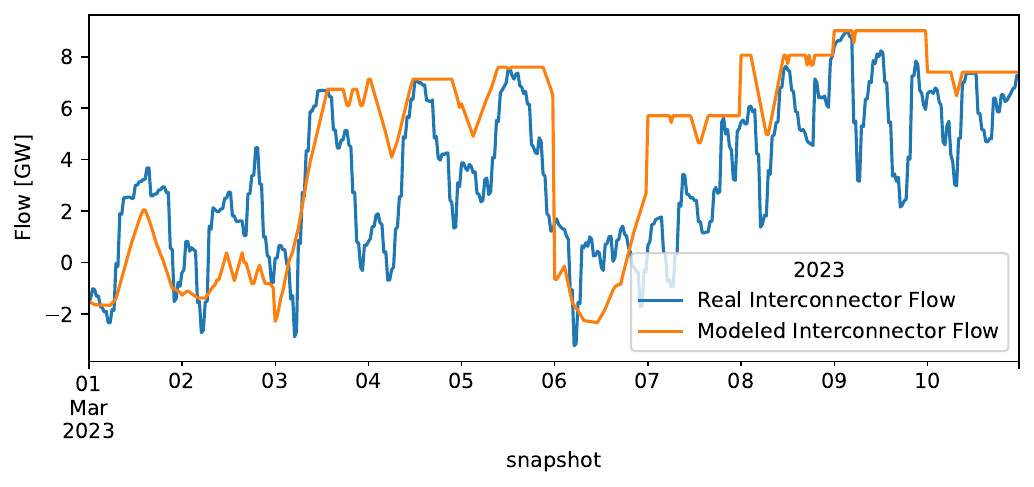} 
  \end{minipage}
  
  \vspace{0.5cm} 
  
  \begin{minipage}[b]{0.49\textwidth}
    \centering
    \includegraphics[width=\textwidth]{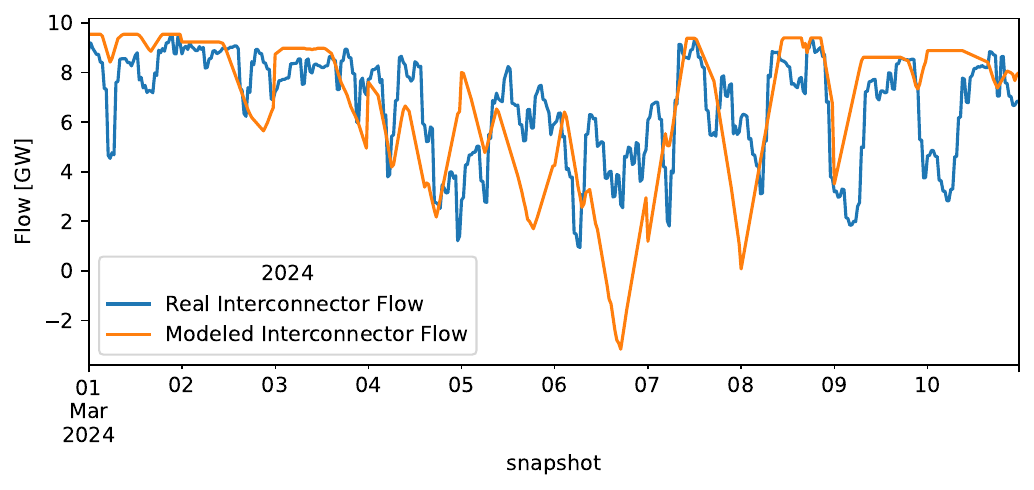} 
  \end{minipage}

  \caption{Comparison of real versus modelled cumulative interconnector flow for the same week in 2022, 2023, and 2024.}
  \label{fig:intercon_comparison}
\end{figure}

\vspace{0.1cm} \paragraph{\textbf{Interconnectors}} \hspace{0.3cm} 
The real behaviour of interconnectors is hard to reproduce, as their operational decisions are often affected by regulatory and grid stability considerations, which are difficult to simulate in linear-optimisation modelling \cite{modo2024intercon}.
Moreover, their operational decisions also depend on accurate simulation of the electricity market in the neighbouring country.
Models that emulate future behaviour therefore often embed their GB model within a market simulation of neighbouring countries \cite{fti2023assessment, franken2025power}.
This, however, is not the approach chosen here.
Instead, the model uses the ENTSO-E API to obtain wholesale electricity prices in neighbouring markets \cite{entsoe2025api} and applies further constraints to simulate some of the aforementioned restrictions on operation.
In particular, these constraints impose upper bounds on the capacity of each interconnector and the rate of change in power flow, based on the respective values observed in the real system. 
With these restrictions, the model approximates real interconnector behaviour reasonably well (Fig.~\ref{fig:intercon_comparison}).
Interconnectors are also assigned a 99\% efficiency to prevent cyclic behaviour.
One improvement for future work could be the introduction of market elasticity in neighbouring markets, whose spot price is currently unresponsive to changes in interconnector flow.
For smaller markets, such as Belgium or the Netherlands, the current approximation - unresponsive prices - could lead to systematic errors.
Estimates of merit order steepness in neighbouring countries exist and could be implemented here \cite{mendes2024euromod}.

\mssubsection[subsec:costs]{Generation Costs}
Transactions on the GB wholesale electricity market are not publicly available, and unit-specific spot prices need to be estimated.
Different generator types require distinct methods, with varying logic and data science tools applied accordingly.

\vspace{0.1cm} \paragraph{\textbf{Nuclear}} \hspace{0.3cm} 
generators serve base load and therefore dispatch continuously.  
This even holds during periods of negative wholesale prices, indicating that nuclear generators are willing to pay to maintain constant dispatch.  
To represent this behaviour, the model iterates over the three-year modelled timeframe and extracts the lowest wholesale price (during which generators were still dispatching) across 2022, 2023, and 2024, and inserts the result as their marginal cost, resulting in £-77.29/MWh.

\vspace{0.1cm} \paragraph{\textbf{Wind with CfD}} \hspace{0.3cm}
Generators with CfDs have their revenue per dispatched energy stabilised to a unit-specific strike price.  
If the wholesale market price is below the strike price, they receive a top-up; if the wholesale price is above the strike price, they return the excess revenue \cite{jansen2020offshore}.  
This mechanism is, however, suspended if there have been more than six hours of continuous negative wholesale prices \cite{cfd2015contract}.  
As a result, it is assumed here that CfD units bid at £0/MWh in the wholesale market.  
The unit's final revenue is then inferred based on actual dispatch in post-processing.  
Unit-level strike prices are made available by the Low Carbon Contracts Company and are used here \cite{lccc2025cfd}.

\vspace{0.1cm} \paragraph{\textbf{Wind under RO scheme}} \hspace{0.3cm}
The remaining wind generators are assumed to receive wholesale revenues and, on top, sell Renewable Obligation Certificates (RO for the policy or ROC for the sold commodity) for a flat price per actually dispatched unit of generation.  
Units differ in the value of their RO, reflecting different levels of technological readiness and the subsidy required for profitable operation.  
The RO scheme is deprecated, and new wind units now receive CfDs instead \cite{ofgem2019ro}.  
Nevertheless, the large majority of wind generators are still on RO schemes.  
Under the assumption that units aim to ensure short-term profits \cite{brown2024price}, the model's RO wind units bid negative their ROC value (Fig.~\ref{fig:merit_order}).

\begin{figure}[h]
    \centering
    \makebox[\textwidth]{\includegraphics[width=0.7\textwidth]{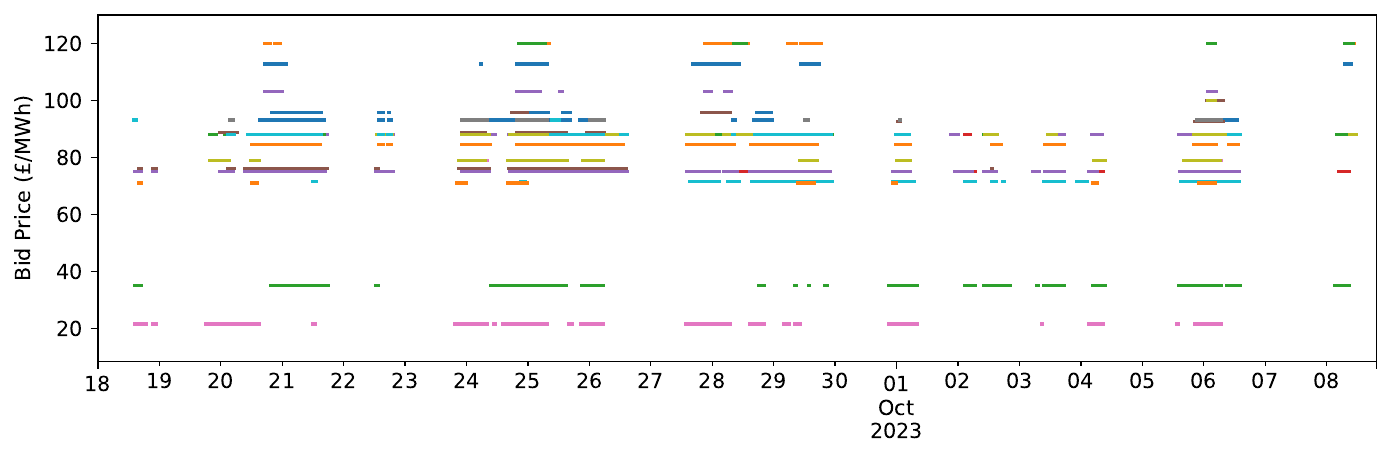}} 
    \caption{Wind generators' bid prices for a subset of 30 generators during September and October 2023.}
    \label{fig:wind_bidding}
\end{figure}

\indent The value of each unit's RO certificate is not publicly available, but can be estimated from the unit's bidding behaviour in the balancing market.  
When a unit curtails, it forgoes revenue from selling ROCs and therefore seeks to recover that revenue on the balancing market instead.  
This behaviour is reflected in wind units’ bidding, which tends to occur consistently at exactly the same unit-specific price (Fig.~\ref{fig:wind_bidding}) - a pattern absent in all other types of assets.  
It is assumed that each unit’s recurring bidding price is the best available approximation of its actual ROC price.  
Only bids that were accepted by the system operator were considered. Therefore, for a subset of units, the ROC is estimated based on the remainder of the cohort.  
The ROC values of those units are sampled from a normal distribution with the mean and variance of the obtained estimates.  
(This introduces bias, as presumably the non-accepted bids are more expensive; future work could evaluate this effect further.)  
Using this method, it is recovered that offshore wind has more expensive ROCs than onshore wind \cite{ofgem2019ro} (Fig.~\ref{fig:merit_order}).

\begin{figure}[h]
    \centering
    \makebox[\textwidth]{\includegraphics[width=0.7\textwidth]{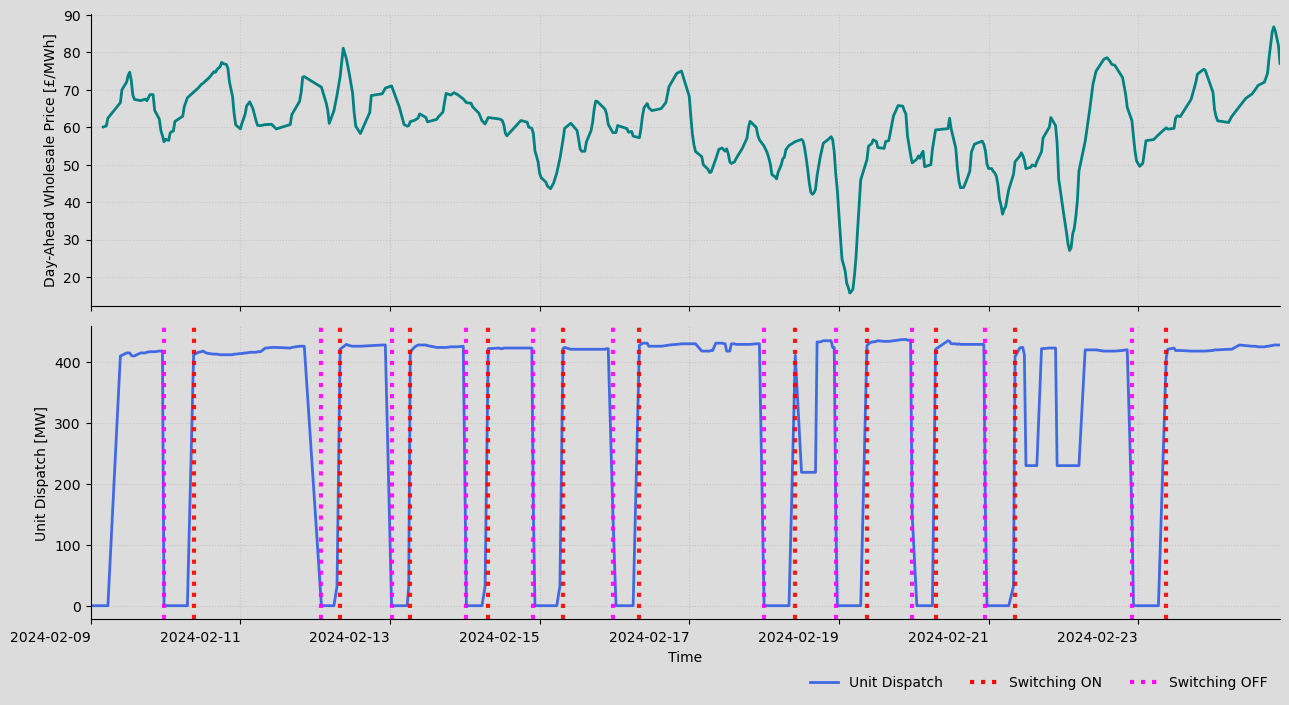}} 
    \caption{Day-ahead price and operation of a thermal unit. The dotted lines indicate the boundaries within which the unit was classified as dispatching, and therefore the day-ahead price was considered when estimating the unit's SRMC.}
    \label{fig:srm_estimate}
\end{figure}

\vspace{0.1cm} \paragraph{\textbf{Thermal Generators}} \hspace{0.3cm}
The short-run marginal cost (SRMC) of gas, coal, biomass, and oil is difficult to estimate.  
The model therefore reverts to a data science approach that considers the dispatch behaviour of each unit and assesses the average day-ahead price during which the generator was willing to dispatch (Fig.~\ref{fig:srm_estimate}).  
This method is applied to every thermal unit in the system.  
However, the varying underlying fuel price is a confounding variable, and therefore, for any modelled day, these values are taken only from the preceding 30 days.  
This method was still unable to recreate the realistic wholesale price fluctuations observed in the real system.  
For that reason, another layer of preprocessing is applied.  

In each settlement period, the model compares the estimated merit order (Fig.~\ref{fig:merit_order}) against the total system load and estimates the price-setting unit.  
(In a multi-period optimisation setting with temporal shifting and interconnectors, this introduces some error.)  
The model then compares that unit's SRMC to the day-ahead price and computes a ``correction factor" from the ratio between them.  
For example, if the day-ahead price is £200/MWh and the national market price-setting thermal unit's SRMC is £100/MWh, that factor would be 2.  
All thermal generators' SRMC values are then multiplied by that factor, resulting in a daily day-ahead price curve that approximates the real one.  
This method addresses the common critique that energy system models produce overly ``flat" merit order curves \cite{mendes2024euromod} by building a stepwise linear merit order curve (Fig.~\ref{fig:merit_order}).

\vspace{0.1cm} \paragraph{\textbf{Dispatch-Only Hydropower}} \hspace{0.3cm}
The model assumes zero marginal cost, except for units that exhibit recurring balancing bid prices, similar to some wind generators (Fig.~\ref{fig:wind_bidding}).  
For those units, bid prices in the wholesale market are set to the negative of the unit's ROC value (which is assumed to be the recurring value in bids).  
There is uncertainty regarding which units are on RO schemes; however, dispatch-only hydropower contributes only a small minority of overall dispatch.

\vspace{0.1cm} \paragraph{\textbf{Pumped-Hydro Storage and Batteries}}
Modelling the operational costs of storage units is beyond the scope of this work.  
For simplicity, the model assumes zero operational costs when charging and discharging.
\hspace{0.3cm}

\mssubsection[subsec:load]{Load}
The total system load is based on the half-hourly Physical Notifications of generators, storages, and interconnectors.  
Following the logic from the ESPENI dataset \cite{wilson2021electrical}, total generation and imports from interconnectors and storages are aggregated.  
Then, electricity exports via interconnectors and storages are subtracted to retain only the share of imports that meets local demand.  
The resulting half-hourly time series of demand is spatially disaggregated based on the load weights estimated in Future Energy Scenarios 2021 \cite{ngeso2021fes} (Fig.~\ref{fig:cap_load_network}).

\begin{figure}[h]
    \centering
    \makebox[\textwidth]{\includegraphics[width=0.33\textwidth]{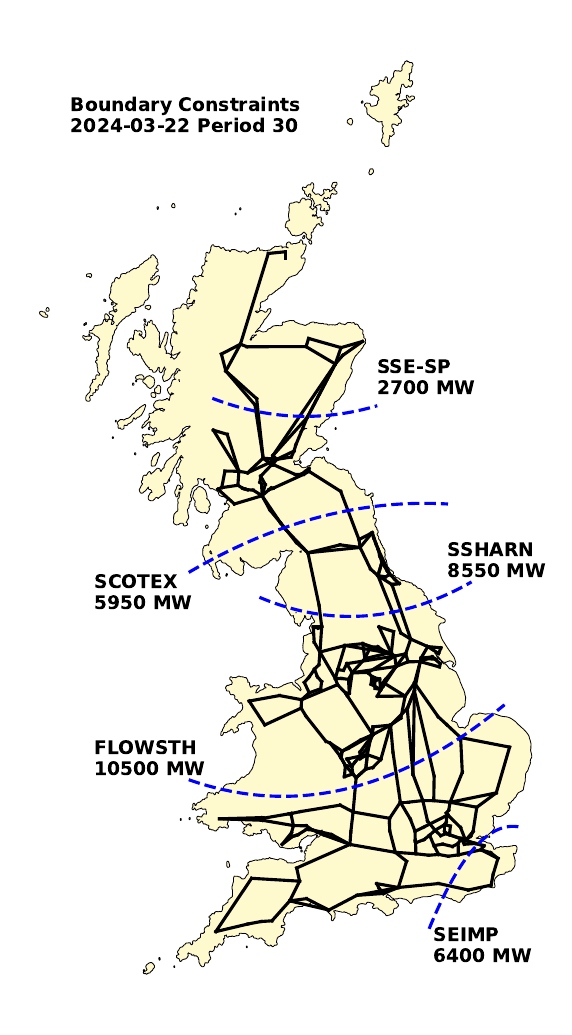}}
    \caption{Transmission grid bottlenecks simulated in the model with thermal constraints for Settlement Period 30 on 2024-03-22.}
    \label{fig:boundaries}
\end{figure}

\mssubsection[subsec:network]{Transmission Network and Interconnectors}
The model uses GB network data from the European-level energy system model PyPSA-Eur \cite{neumann2023potential}, which contains around 300 buses representing Grid Supply Points and the respective power lines connecting them.  
However, a faithful recreation of real power flows is beyond the scope of this work, and the model therefore focuses instead on system behaviours that are crucial for accurately simulating a counterfactual locational market.  
To this end, PyPSA \texttt{Line} components are replaced with the \texttt{Link} component, which forgoes constraining power flow according to Kirchhoff's Voltage Law \cite{brown2017pypsa}, similar to the approach taken in \cite{fti2023assessment}.  
The model next considers day-ahead thermal constraints across five critical transmission boundaries (Fig.~\ref{fig:boundaries}), with the respective data drawn from the National Energy System Operator (NESO).  
From north to south, these are SSE-SP, SCOTEX (B6), SSHARN, FLOWSTH, and SEIMP.  
The capacity of lines (now links) crossing a boundary is scaled such that the total permissible flow across the boundary is given by the thermal constraint (expressed as Net Transfer Capacity).  
The same scaling factor is applied to the lines (now links) surrounding the boundary to avoid creating an unnatural capacity gap between boundary and non-boundary lines (now links).

\indent It is into this network that components and loads are inserted.  
For the national wholesale market, all buses are aggregated into one.  
For the zonal wholesale market, the capacities of lines (now links) that connect two nodes within the same zone are set to infinity, as is done in \cite{fti2023assessment}.

\indent However, the model applies one further step of data-driven line (now link) capacity tuning.  
First, it simulates the national wholesale market schedule, inserts the resulting positions of storage units and interconnectors into the nodal layout, and then re-optimises dispatch (for units that are not storages or interconnectors) again - this time under the consideration of transmission constraints.  
The difference in dispatch between the two optimisations reflects the constraint-management-related share of balancing operations.  
In an iterative approach, the line (now link) capacities are tuned uniformly until the modelled balancing volume matches the real grid-congestion-related balancing volume in the real system for that day.  
The result is a global ``line tuning factor".  
That factor is then applied to the zonal model before its optimisation (and the respective balancing run in the nodal layout).  
Therefore, in effect, the aforementioned thermal constraints across the five boundaries only serve to adjust their relative magnitudes correctly - the driving force in adjusting grid constraints is the attempt to match real and modelled balancing volumes. 

\indent Two ways in which future work could improve on this are by inserting the more recent and improved network model based on OpenStreetMap data \cite{xiong2025modelling}, which is already included in the European-level model, or by replicating similar methods in a linearised AC power flow system.

\mssubsection[subsec:balancing]{Balancing Market}
In times without grid congestion, national and locational markets result in the same wholesale schedule - differences in system behaviour occur only when the grid is congested.  
To match the frequency of these occurrences between the modelled and real systems, as discussed in Methods \ref{subsec:network}, line capacities are calibrated each day such that real daily balancing volumes are matched.  
In the model, balancing volume in the national or zonal layout is defined as the difference between the optimised (national or zonal) wholesale schedule and the nodal (transmission-grid-compliant) actual dispatch.  
Due to degeneracy in marginal prices, this comparison is not conducted at the unit level but instead groups generators into categories - such as wind units north of SCOTEX (B6) or thermal generators south of SCOTEX.
Differences in the aggregate schedule versus dispatch of these groups are considered grid congestion that requires balancing actions. 
The costs assigned to these actions are the costs of real congestion-managing balancing actions for that day, retrieved from Elexon \cite{elexon}.  
In particular, all daily accepted balancing bids and offers are pooled and assigned to the total volume of balancing actions observed in the model.  
This approach falls short of explicitly modelling the balancing market - a limitation discussed in Subsection \ref{subsec:limitations}.  
Technically, explicit modelling is feasible since unit-level price and volume data for bids and offers are available from Elexon \cite{elexon}.  
Macroscopically, however, the central dynamics of the GB balancing market are captured by the existing approach.

\mssubsection[subsec:seb]{Socioeconomic Benefit/Welfare Gain}
The socioeconomic benefit (SEB) is defined as a reduction in the operational cost of the GB power system while meeting the same load.  
A zonal market can achieve SEB through two main dynamics: first, by enabling more domestic wind generation through zonal price signals; and second, by visualising local electricity scarcity, thereby improving interconnector and battery operation.  
These dynamics manifest as five components centred around fuel savings and improved operation, which are summed here to derive the total SEB.  
The methodology follows \cite{fti2023assessment}.

\textit{1) Export revenues.} The change in exports across interconnectors is multiplied by GB's wholesale price.

\textit{2) Import cost.} The change in imports across interconnectors is multiplied by the neighbouring country's wholesale price.

\textit{3) Interconnector congestion rents.} Power flow and the wholesale price difference across interconnectors are tracked, and 50\% of the change is attributed to GB.

\textit{4) Prevented Thermal Balancing Short-Run Marginal Cost (SRMC).} This work considers the full avoided cost of operating a thermal unit that is active in the national, but not the zonal, dispatch as a socio-economic benefit.  
If the prevented service was procured in the balancing market, the prevented cost is the wholesale marginal cost of the unit during those settlement periods added to a balancing markup, which throughout the paper is set to £30/MWh, a value aligned with \cite{entsoe2025bzr}.
For an overview of how units' marginal cost is estimated see Methods \ref{subsec:costs}.

\textit{5) Prevented Thermal Wholesale SRMC.} If the prevented service was procured in the wholesale market, the prevented cost is the wholesale marginal cost of the unit during those settlement periods.



\mssubsection[subsec:limitations]{Limitations}
The primary source of inaccuracy is likely to be the approximation of power flow, followed by the uncertainty in estimates made during the modelling of the merit-order curve. \\
\indent With regard to power flow, the model applies a DC approximation, which neglects Kirchhoff's Law.  
This method was also used in reports such as \cite{fti2023assessment}, and has here been further enhanced through data-driven network capacity calibration (Methods \ref{subsec:network}).  
The main advantage of the DC formulation is that it yields a straightforward zonal market definition (Methods \ref{subsec:markets}).  
For this reason, the chosen method appears to be appropriate.  
However, internal experiments have shown that incorporating Kirchhoff's Voltage Law alters the spatial distribution of marginal prices, usually resulting in a more complex layout rather than a binary structure with one high- and one low-price region.  
During that experiment, the zonal market layout was simulated using the \url{cluster_network.py} script provided by PyPSA-Eur \cite{horsch2018pypsa}, in contrast to the method used in this study (see Methods \ref{subsec:markets}).  
The network aggregation executed by \url{cluster_network.py} follows a logic not specifically designed for locational pricing studies.  
In particular, the script balances transmission capacity across regional boundaries against the intra-regional ability of the grid to transport electricity from supply to demand.  
It is therefore a less accurate representation of the zonal cross-border transmission capacities that a grid operator would consider.  
Future work should improve \textit{GBPower} by implementing inter-zonal market coupling that more accurately approximates the complexity of the real system, inspired, for instance, by \cite{savelli2017optimization}.

\indent To estimate the marginal generation costs of thermal units, the model opts for a data-based approach, correlating day-ahead prices with each unit's dispatch and then, for each settlement period, scaling the marginal price of the price-setting unit such that it matches the real day-ahead price (Methods \ref{subsec:costs}).  
However, this approach neglects both the engineered properties that are likely to primarily drive plant dispatch costs, and the unit-commitment nature of the optimisation problem.  
For the sake of computational feasibility, the model uses a strictly linear optimisation formulation, and therefore fails to simulate more nuanced decision-making by plant operators.  
Nonetheless, the results are likely to be a reasonable approximation of the real system, since the number of included units is relatively large, and the collective behaviour of a group of unit-committing plants can approximate a linear formulation of the problem.

\indent Modelling the balancing market explicitly is feasible, since Elexon \cite{elexon} shares not only accepted but also all submitted bids and offers.  
Bid and offer prices and volumes could be assigned to units, and the nodal optimisation model could determine independently which ones are cost-optimal to accept.  
This has not been implemented in the current version of the model but could provide insights into market liquidity or opportunities for strategic bidding.  
However, for the present assessment, different dynamics are being analysed, and a simpler treatment appears justified.

\indent Each model run only uses a 24-hour optimisation horizon.  
This may result in errors for a subset of storage assets, in particular pumped-hydro storage, where an analysis of the respective $E/P$ ratios suggests that inter-day optimisation could be feasible.  
However, this limitation is accepted in favour of a more concise and largely accurate day-by-day framing.

%% file: sections/appendix.tex
\begin{figure}[h]
    \centering
    \makebox[\textwidth]{\includegraphics[width=1.1\textwidth]{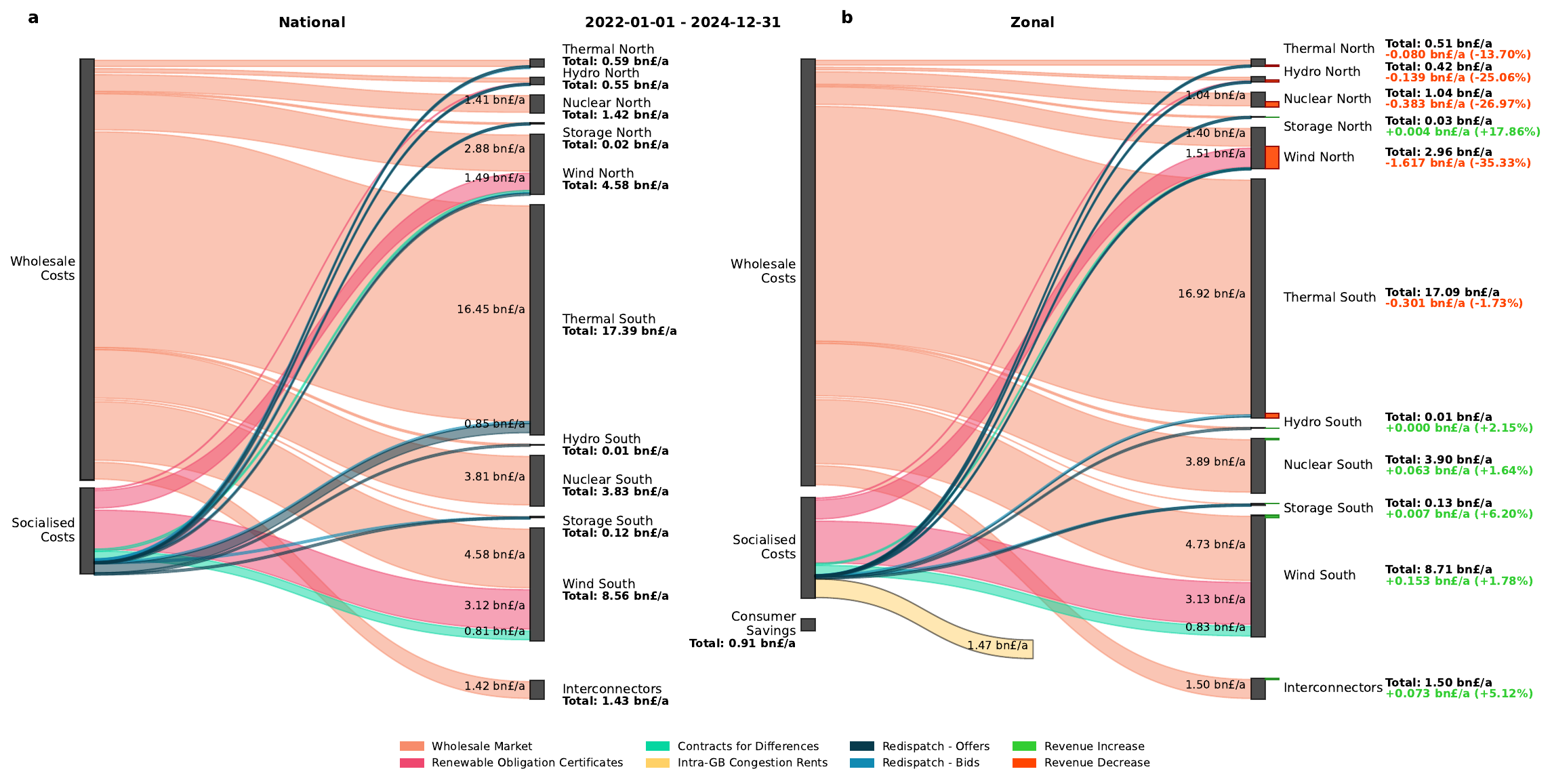}}
    \caption{\textbf{Change in payment flows from consumers to producers for national and zonal markets.}  
    \textit{Socialised Costs} refers to all payments levied based on load. The grouping into northern and southern assets is based on Figure \ref{fig:north_south_split}.}
    \label{fig:sankey}
\end{figure}

\begin{figure}[h]
    \centering
    \makebox[\textwidth]{\includegraphics[width=1.\textwidth]{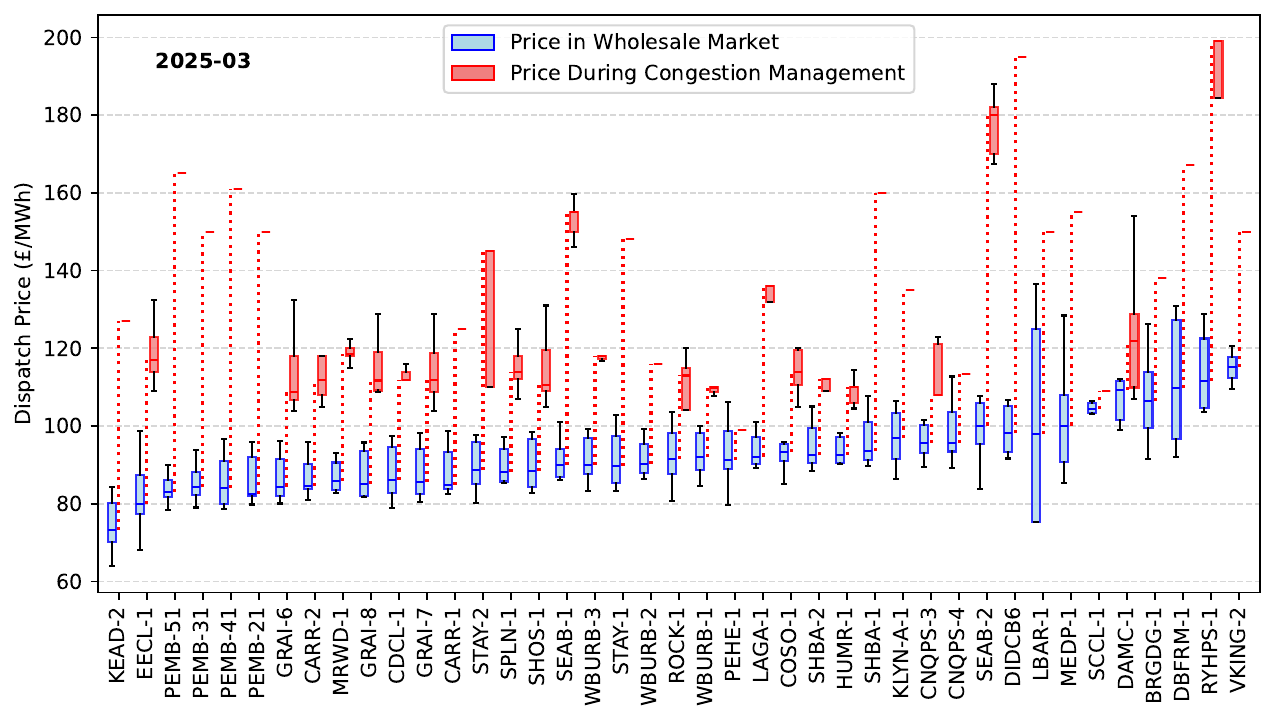}}
    \caption{\textbf{Comparison of dispatch prices when the same units dispatch in wholesale versus balancing markets for March 2025.}  
    Wholesale market dispatch prices are estimated based on Methods \ref{subsec:costs}; redispatch prices are the unit's accepted offers during the month.}
    \label{fig:dispatch_price}
\end{figure}

\begin{figure}[h]
    \centering
    \makebox[\textwidth]{\includegraphics[width=1.2\textwidth]{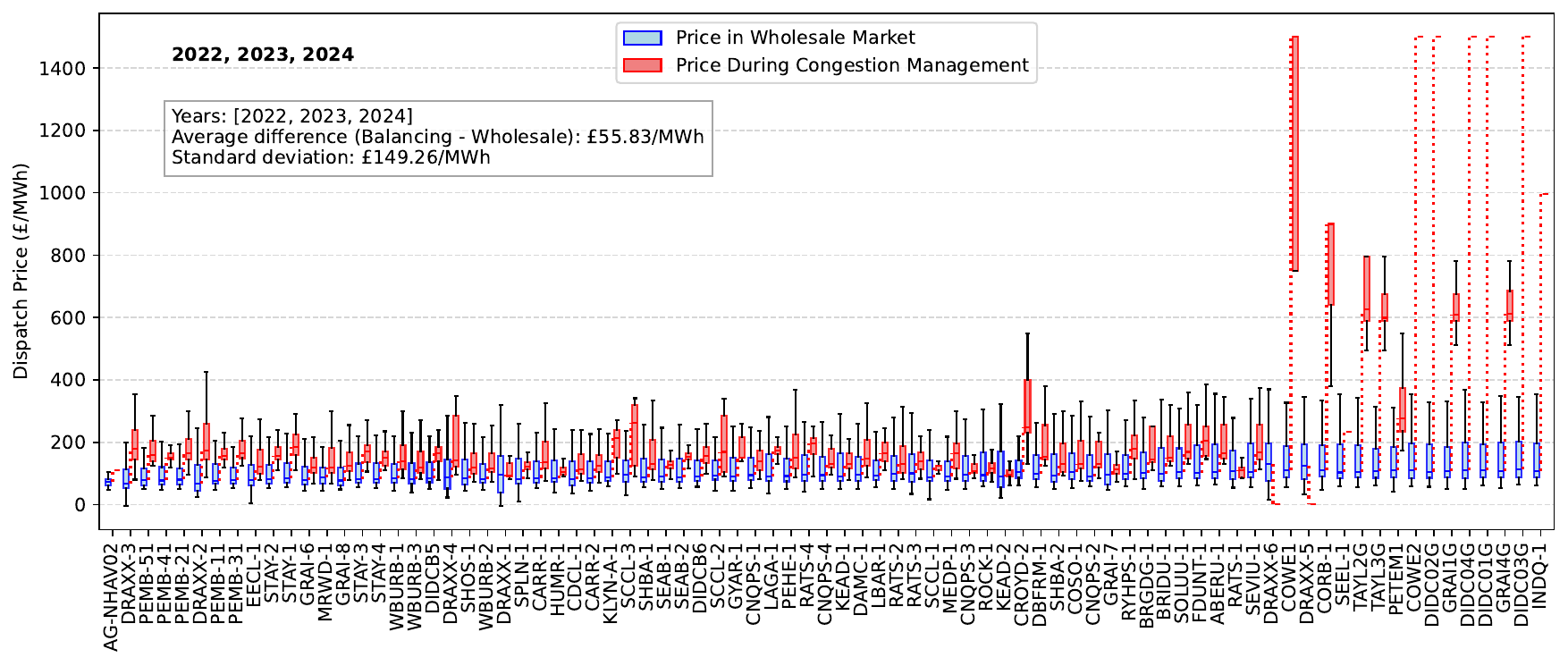}}
    \caption{\textbf{Comparison of dispatch prices when the same units dispatch in wholesale versus balancing markets for 2022, 2023, and 2024.}  
    Wholesale market dispatch prices are estimated based on Methods \ref{subsec:costs}; redispatch prices are the unit's accepted offers over the same timeframe.}
    \label{fig:dispatch_price_years}
\end{figure}

\begin{figure}[h]
    \centering
    \makebox[\textwidth]{\includegraphics[width=1.\textwidth]{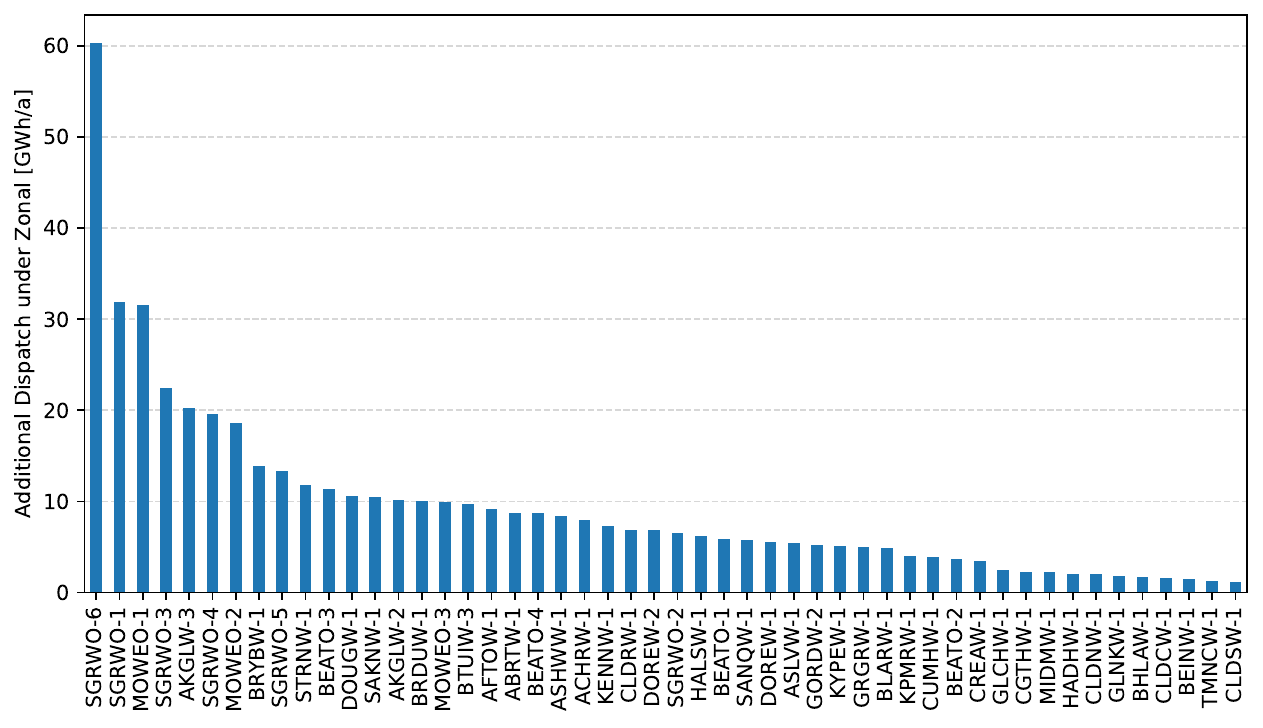}} 
    \caption{\textbf{Average annual increase of actual dispatch of different wind generators in a zonal market.} This is not day-ahead but actual dispatch, totalling for three years to 1.37TWh.}
    \label{fig:dispatch_increase_gwh}
\end{figure}

\begin{figure}[h]
    \centering
    \makebox[\textwidth]{\includegraphics[width=1.\textwidth]{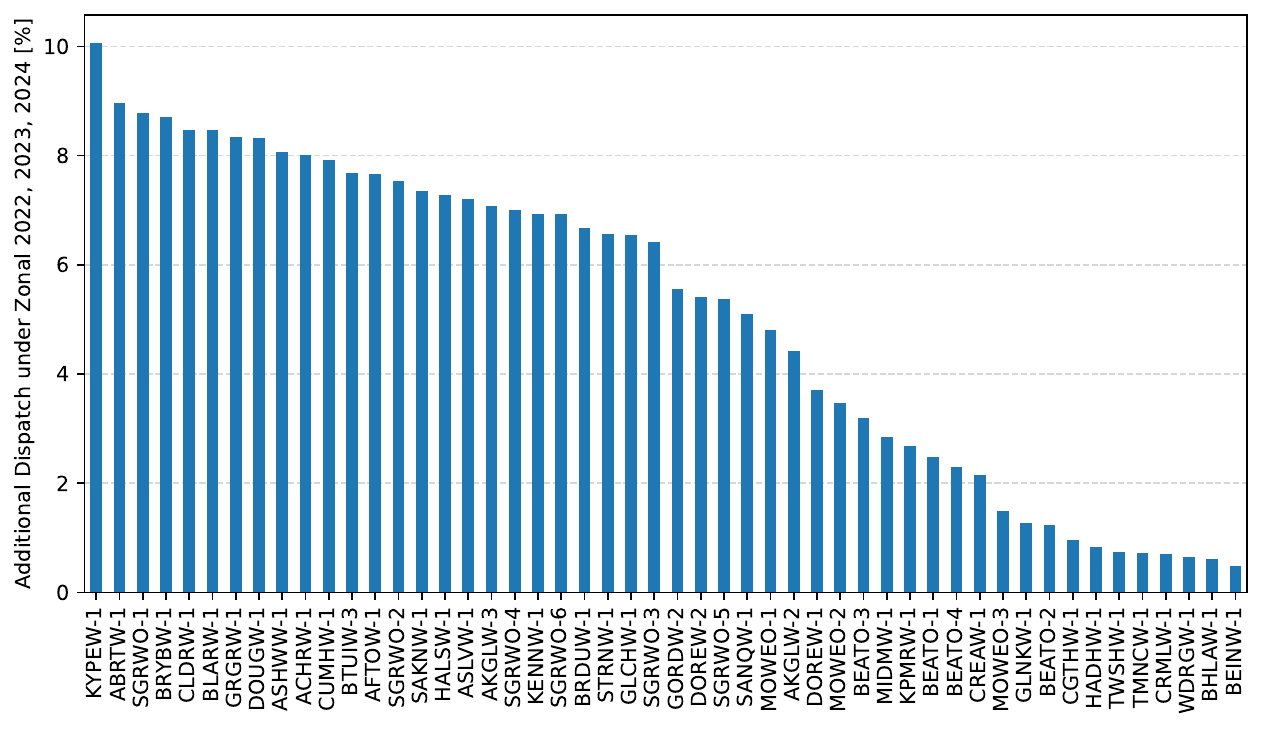}} 
    \caption{\textbf{Average annual percentage increase of actual dispatch of different wind generators in a zonal market.}}
    \label{fig:dispatch_increase_perc}
\end{figure}

\begin{figure}[h]
    \centering
    \makebox[\textwidth]{\includegraphics[width=\textwidth]{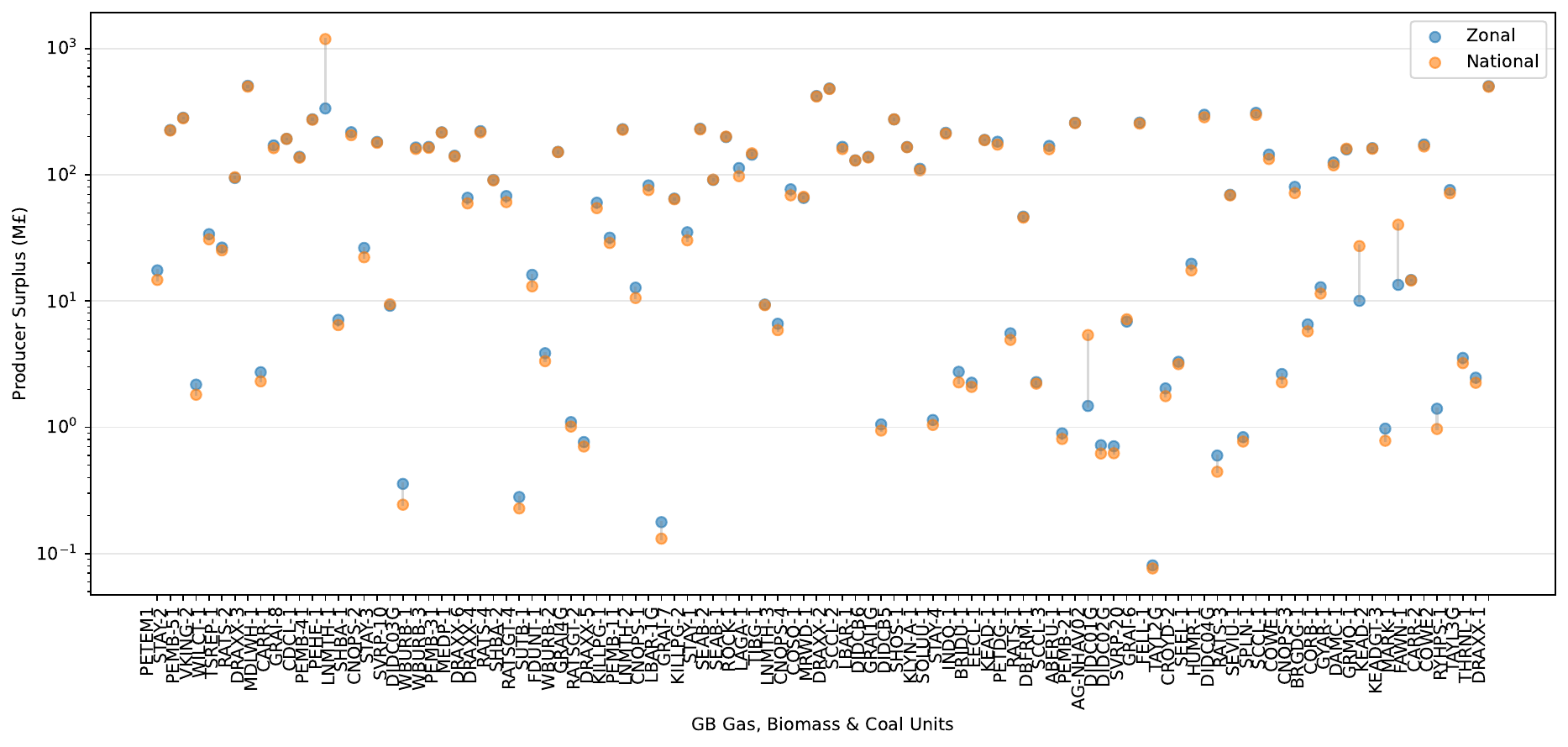}} 
    \caption{Absolute thermal generator surplus over three year period.}
    \label{fig:thermal_schedule_revenue}
\end{figure}

\begin{figure}[h]
    \centering
    \makebox[\textwidth]{\includegraphics[width=\textwidth]{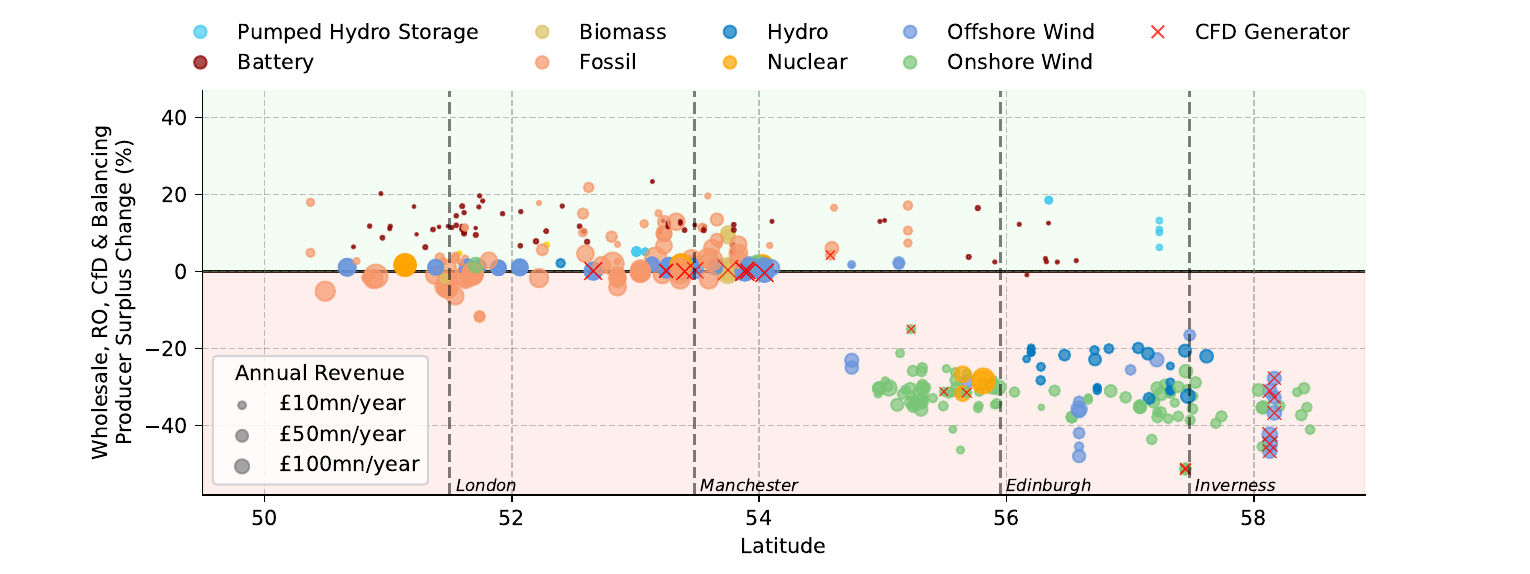}} 
    \caption{Percentage producer surplus change, assuming that thermal units receive a premium of £15/MWh for balancing.}
    \label{fig:surplus_15}
\end{figure}

\begin{figure}[h]
    \centering
    \makebox[\textwidth]{\includegraphics[width=1.\textwidth]{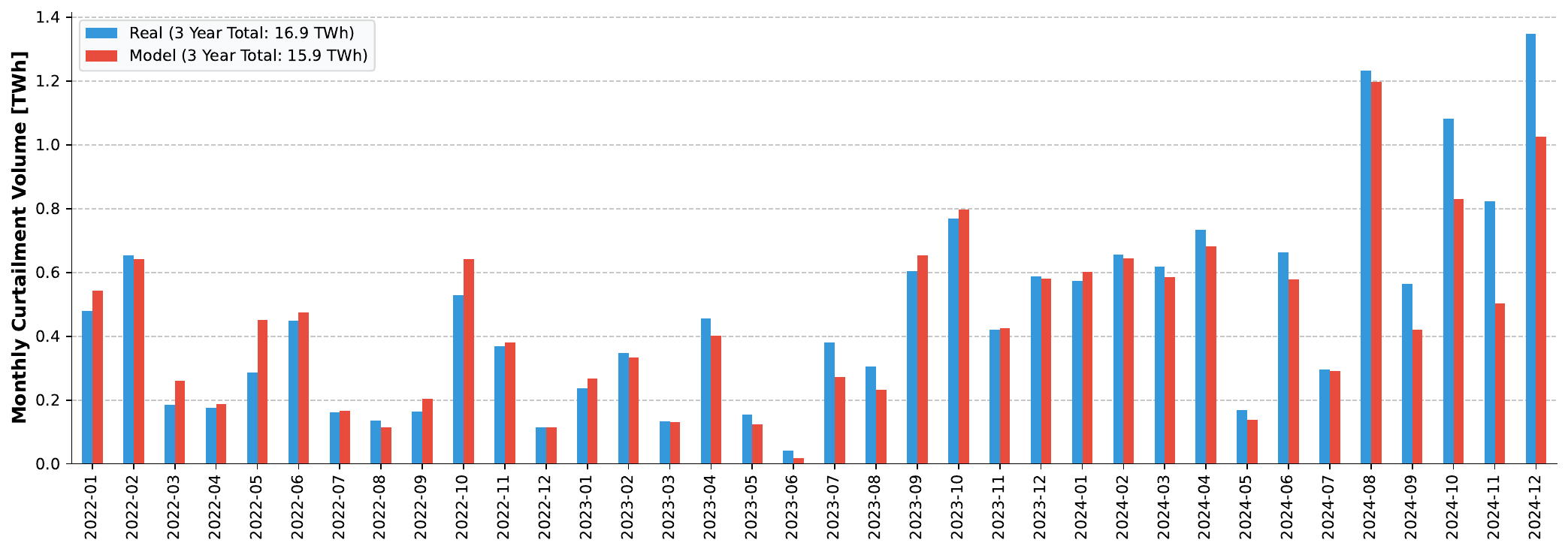}} 
    \caption{\textbf{Comparison between real grid-congestion-related curtailment volume and modelled volumes.}  
All accepted bids in Elexon \cite{elexon} that are tagged with the \texttt{soFlag} are interpreted as grid-congestion-related curtailment.
}
    \label{fig:balancing_volume_validation}
\end{figure}

\begin{figure}[h]
    \centering
    \makebox[\textwidth]{\includegraphics[width=0.5\textwidth]{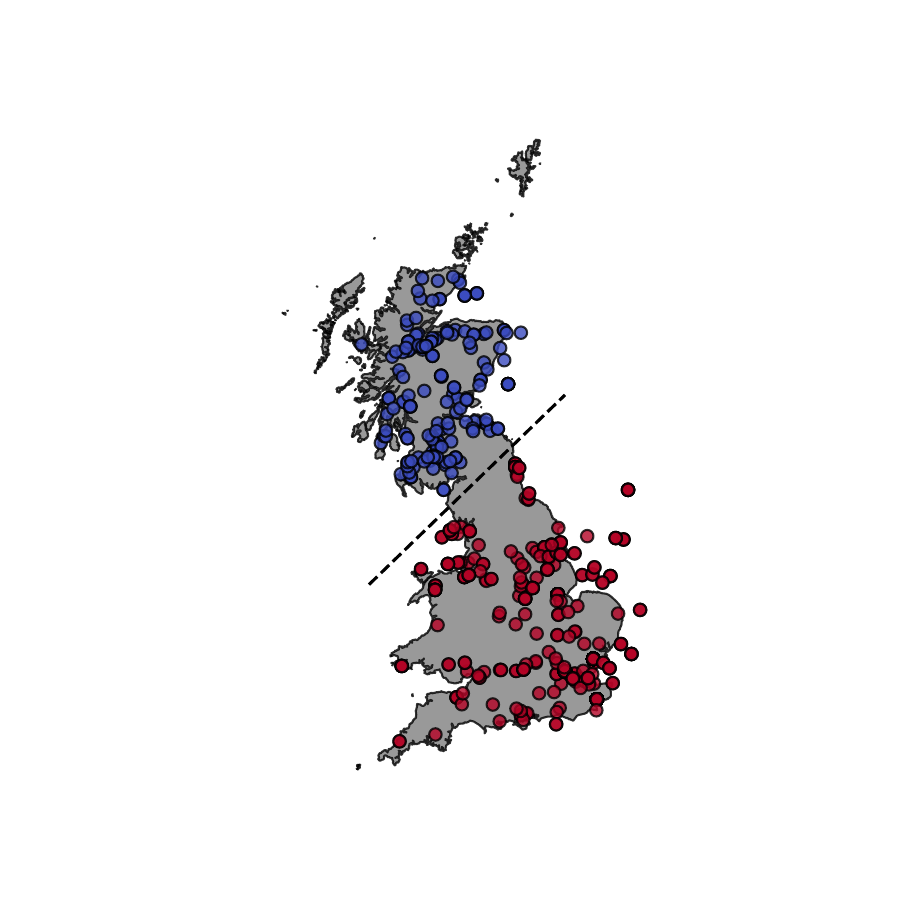}} 
    \caption{North-South classification of system assets.}
    \label{fig:north_south_split}
\end{figure}

